\documentclass[letterpaper,twocolumn,10pt]{article}
\usepackage{usenix2019_v3}

\usepackage{tikz}
\usepackage{amsmath}
\usepackage{titling}

\usepackage{algorithm}
\usepackage{amsthm}
\usepackage{amssymb}
\usepackage{multirow}
\usepackage{bigstrut}
\usepackage{comment}
\usepackage[noend]{algpseudocode}

\usepackage{ifthen}

\usepackage[T1]{fontenc}
\usepackage{inconsolata}

\usepackage{graphicx}
\usepackage{subcaption}
\usepackage{textcomp}
\usepackage{tabularx}
\usepackage{enumitem}
\usepackage{cite}
\usepackage{listings}
\usepackage{filecontents}
\usepackage{xcolor}
\usepackage{multirow}
\usepackage{xspace}
\usepackage{url}
\usepackage[binary-units]{siunitx}

\makeatletter
\def\do@url@hyp{\do\-\do\_}
\makeatother

\definecolor{mygreen}{rgb}{0,0.6,0}
\definecolor{mygray}{rgb}{0.5,0.5,0.5}
\definecolor{mymauve}{rgb}{0.58,0,0.82}
\definecolor{myred}{rgb}{0.79,0.15,0.15}

\lstdefinestyle{mypython}{
  language=Python,
  backgroundcolor=\color{white},   
  basicstyle=\scriptsize\ttfamily,        
  breakatwhitespace=false,         
  breaklines=true,                 
  captionpos=b,                    
  commentstyle=\color{mygreen},
  deletekeywords={...},            
  escapeinside={\%*}{*)},          
  extendedchars=true,              
  firstnumber=1,                
  frame=no,                    
  keepspaces=true,                 
  keywordstyle=\color{mymauve},       
  language=Octave,                 
  morekeywords={DIRECT, PERIODIC, IMMEDIATE, BY_TIME, EVERY_OBJ},            
  numbers=left,                    
  numbersep=5pt,                   
  numberstyle=\tiny\color{mygray}, 
  rulecolor=\color{black},         
  showspaces=false,                
  showstringspaces=false,          
  showtabs=false,                  
  stepnumber=1,                    
  stringstyle=\color{myred},     
  tabsize=4,                    
  title=\lstname                   
}

\lstdefinestyle{trigger}{
  language=C++,
  basicstyle=\scriptsize\ttfamily,
  breaklines=true,
  commentstyle=\color{mygreen},
  captionpos=b,
  keywordstyle=\color{blue},
  morekeywords={abstract, string},
  showspaces=false,  
  showstringspaces=false,          
  showtabs=false,  
  tabsize=2,                    
  title=\lstname  
}

\lstdefinestyle{mycpp}{
  language=C++,
  basicstyle=\scriptsize\ttfamily,
  breaklines=true,
  commentstyle=\color{mygreen},
  keywordstyle=\color{blue},
  firstnumber=1,   
  numbers=left,
  numbersep=5pt,
  captionpos=b,
  showspaces=false,  
  numberstyle=\tiny\color{mygray},
  showstringspaces=false,          
  stringstyle=\color{myred},     
  showtabs=false,  
  tabsize=2,                    
  title=\lstname  
}

\lstdefinestyle{interface}{
  language=C++,
  basicstyle=\footnotesize\ttfamily,
  breaklines=true,
  keywordstyle=\color{blue},
  captionpos=b,                    
}

\newfloat{lstfloat}{htbp}{lop}  
\floatname{lstfloat}{Listing}

\newcommand{\PHB}[1]{\noindent\textbf{#1}\hspace{.5em}} 
\newcommand{\PHM}[1]{\vspace{.2em}
\noindent\textbf{#1}\hspace{.5em}} 

\newcommand{\SysName}{\texttt{Pheromone}\xspace}
\newcommand{\SysNameMR}{\texttt{Pheromone-MR}\xspace}

\newcommand{\final}[1]{{#1}}
\newcommand{\revise}[1]{{#1}}
\newcommand{\new}[1]{{#1}}

\newboolean{exclude_appendix}   
\setboolean{exclude_appendix}{false}

\graphicspath{{figures/}}

\newcommand{\ww}[1]{\textcolor{red}{\textbf{[ww: #1]}}}
\posttitle{\par\end{center}}

\begin{document}

\date{}

\title{\Large \bf Following the Data, Not the Function:\\Rethinking Function Orchestration in Serverless Computing}





\author{
	{\rm Minchen Yu$^{\dag}$ \quad Tingjia Cao$^{\ddag}$\thanks{This work was partially done while the author was at HKUST.} \quad Wei Wang$^{\dag}$ \quad Ruichuan Chen$^{\S}$}\\
	$^{\dag}$Hong Kong University of Science and Technology\\
	$^{\ddag}$University of Wisconsin-Madison \quad $^{\S}$Nokia Bell Labs
} 


\maketitle
\begin{abstract}

    Serverless applications are typically composed of
    function workflows in which multiple short-lived functions are
    triggered to exchange data in response to events or state changes.
    Current serverless platforms coordinate and trigger functions by
    following high-level invocation dependencies but are oblivious to the
    underlying data exchanges between functions. This design is
    neither efficient nor easy to use in orchestrating complex workflows --
    developers often have to manage complex function interactions by themselves,
    with customized implementation and unsatisfactory performance.



    In this paper, we argue that function orchestration should follow a
    \emph{data-centric approach}. In our design, the platform provides a
    \emph{data bucket abstraction} to hold the intermediate data generated
     by functions. Developers can use a rich set of data trigger primitives to
     control when and how the output of each function
     should be passed to the next functions in a workflow. 
     By making data consumption explicit and allowing it to
     trigger functions and drive the workflow, complex function interactions
     can be easily and efficiently supported. We present \SysName{} -- a
     scalable, low-latency serverless platform following this data-centric
     design. Compared to well-established commercial and open-source platforms, \SysName
     cuts the latencies of function interactions and data exchanges by orders
     of magnitude, scales to large workflows, and enables easy
     implementation of complex applications.

\end{abstract}

\section{Introduction}
\label{sec:intro}

Serverless computing, with its Function-as-a-Service incarnation, is becoming
increasingly popular in the cloud. It allows developers to write highly
scalable, event-driven applications as a set of short-running
functions. Developers simply specify the events that trigger the activation
of these functions, and let the serverless platform handle resource provisioning,
autoscaling, logging, fault-tolerance, etc. Serverless computing is also
economically appealing as it has zero idling cost: developers are only
charged when their functions are running.


Many applications have recently been migrated to the severless cloud~\cite
{lambda_scenarios, jonas_occupy_2017,pu_shuffling_2019,carver_wukong_2020,
fouladi_encoding_nodate,ao_sprocket_2018,yu_gillis_icdcs,zhang_mark:_2019,tian2022Owl}.
These applications typically consist of multiple interactive functions with
diverse function-invocation and data-exchange patterns. For example, a
serverless-based batch analytics application may trigger hundreds of parallel
functions for all-to-all data communication in a shuffle phase~\cite
{pu_shuffling_2019, klimovic_pocket:_2018, zhang_caerus_nodate}; a stream
processing application may repeatedly trigger certain functions to process
dynamic data received in a recent time window. Ideally, a serverless
platform should provide an \emph{expressive} and \emph{easy-to-use} function
orchestration to support various function-invocation and
data-exchange patterns. The orchestration should also be made \emph
{efficient}, enabling low-latency invocation and fast data exchange between
functions.



However, function orchestration in current serverless platforms is neither efficient nor
easy to use. It typically models a serverless application as a \emph{workflow} that
connects functions according to their invocation dependency~\cite
{akkus_sand:_2018, knix,sreekanti_cloudburst_2020, kotni_faastlane_nodate,
mahgoub_sonic_nodate,open_whisk_composer,google_cloud_composer,aws_step_function}.
It specifies the order of function invocations but is \emph{oblivious to
when and how data are exchanged between functions}. Without such
knowledge, the serverless platform assumes that the output of a function is
entirely and immediately consumed by the next function(s), which is not the
case in many applications such as the aforementioned ``shuffle'' operation in batch
analytics and the processing of dynamically accumulated data in stream
analytics. To work around these limitations, developers have to manage
complex function interactions and data exchanges by themselves, using various
approaches such as a message broker or a shared storage,
either synchronously or asynchronously~\cite
{sreekanti_cloudburst_2020,
aws_cache,aws_s3,knix,klimovic_pocket:_2018,step_func_large_data}. As no
single approach is found optimal in all scenarios, developers may need to write
complex logic to dynamically select the most efficient approach at runtime
(see \S\ref{sec:deploy_app}). Current serverless platforms also incur function
interaction latencies of tens of milliseconds, which may be unacceptable to
latency-sensitive applications~\cite{jia_nightcore_2021}, particularly since
this overhead accumulates as the function chain builds up.

In this paper, we argue that function orchestration should follow the flow of
data rather than the function-level invocation dependency, thus 
a \emph{data-centric approach}. Our key idea is to make data consumption
explicit, and let it trigger functions and drive the workflow. In our design,
the serverless platform exposes a \emph{data bucket abstraction} that holds the
intermediate output of functions in a logical object store. The data bucket
provides a rich set of data trigger primitives that developers can use to
specify \emph{when} and \emph{how} the intermediate data are passed to the
intended function(s) and trigger their execution. With such a fine control of
data flow, developers can express sophisticated function invocations and data
exchanges, simply by configuring data triggers through a unified interface.
Knowing how intermediate data will be consumed also enables the serverless
platform to schedule the intended downstream functions close to the input,
thus ensuring fast data exchange and low-latency function invocation.


Following this design approach, we develop \SysName\footnote{Pheromone is a
chemical signal produced and released into the environment by an animal that
triggers a social response of others of its species. We use it as a metaphor
for our data-centric function orchestration approach.}, a scalable serverless
platform with low-latency data-centric function orchestration. \SysName
proposes three key designs to deliver high performance. \textbf{First}, it uses
a two-tier distributed scheduling hierarchy to locally execute a function
workflow whenever possible. Each worker node runs a local
scheduler, which keeps track of the execution status of a workflow via its data buckets and
schedules next functions of the workflow onto local function executors.
In case that all local executors are busy, the scheduler forwards the request
to a global coordinator which then routes it to another worker node with
available resources. 
\textbf{Second}, \SysName trades the durability of
intermediate data, which are typically short-lived and immutable, for fast
data exchange. Functions exchange data within a node through a zero-copy
shared-memory object store; they can also pass data to a remote function
through direct data transfer.
\textbf{Third}, \SysName uses sharded global coordinators,
each handling a disjoint set of workflows. With such a shared-nothing
design, local schedulers only synchronize workflows' execution
status with the corresponding global coordinators, which themselves require
no synchronization, thus ensuring high scalability for distributed
scheduling.


We evaluate \SysName against well-established commercial and open-source
serverless platforms, including AWS Lambda with Step Functions, Azure
Durable Functions, Cloudburst~\cite{sreekanti_cloudburst_2020}, and
KNIX~\cite{knix}. Evaluation results show that \SysName improves the function
invocation latency by up to 10$\times$ and 450$\times$ over Cloudburst
(best open-source baseline) and AWS Step Functions (best commercial
baseline), respectively. \SysName scales well to large workflows and incurs only
millisecond-scale orchestration overhead when running 1k chained functions
and 4k parallel functions, whereas the overhead is at least a few seconds in
other platforms. \SysName has negligible data-exchange overhead (e.g., tens
of $\mu$s), thanks to its zero-copy data exchange. It can also handle failed
functions through efficient re-execution. Case studies of two
serverless applications, i.e., Yahoo! stream processing~\cite
{chintapalli_benchmarking_2016} and MapReduce sort, further demonstrate
that \SysName can easily express complex function interaction patterns (\emph
{rich expressiveness}), require no specific implementation to handle data
exchange between functions (\emph{high usability}), and efficiently support
both latency-sensitive and data-intensive applications (\emph
{wide applicability}).

\section{Background and Motivation}
\label{sec:motivation}

We first introduce serverless computing and discuss the
limitations of function orchestration in current serverless platforms.

\subsection{Serverless Computing}
\label{sec:serverless_background}

Serverless computing, with its popular incarnation being
Function-as-a-Service (FaaS), has recently emerged as a popular cloud
computing paradigm that supports highly-scalable, event-driven
applications~\cite{aws_lambda,azurefunc,google_cloud_function}. Serverless
computing allows developers to write short-lived, stateless functions that
can be triggered by events. The interactions
between functions are simply specified as \emph{workflows}, and the
serverless platform manages resource provisioning, function orchestration,
autoscaling, logging, and fault tolerance for these workflows. This paradigm
appeals to many developers as it allows them to concentrate on the
application logic without having to manage server resources~\cite
{jonas2019cloud,Schleier-Smith2021a} -- hence the name serverless computing.
In addition to the high scalability and operational simplicity, serverless
computing adopts a ``pay-as-you-go'' billing model: developers are billed only when their
functions are invoked, and the function run-time is metered at a fine
granularity, e.g., 1~ms in major serverless platforms~\cite
{aws_lambda,azurefunc}. Altogether, these benefits have increasingly driven a large
number of traditional ``serverful'' applications to be migrated to the serverless
platforms, including batch analytics~\cite
{jonas_occupy_2017,pu_shuffling_2019,carver_wukong_2020,zhang_caerus_nodate},
video processing~\cite{ao_sprocket_2018,fouladi_encoding_nodate}, stream
processing~\cite{lambda_scenarios}, machine learning~\cite
{yu_gillis_icdcs,zhang_mark:_2019}, microservices~\cite
{jia_nightcore_2021}, etc.


\subsection{Limitations of Current Platforms}
\label{sec:deploy_app}

Current serverless platforms take a \emph{function-oriented} approach to
orchestrating and activating the functions of a serverless workflow: each
function is treated as a single and standalone unit, and the interactions of
functions are separately expressed within a workflow. This workflow connects
individual functions according to their invocation dependencies, such that
each function can be triggered upon the completion of one or multiple
upstream functions. For example, many platforms model a serverless workflow
as a directed acyclic graph (DAG)~\cite
{akkus_sand:_2018, knix,sreekanti_cloudburst_2020, kotni_faastlane_nodate,
mahgoub_sonic_nodate,
open_whisk_composer,google_cloud_composer,aws_step_function}, in which the
nodes represent functions and the edges indicate the invocation dependencies
between functions. The DAG can be specified using general programming
languages~\cite{open_whisk_composer,google_cloud_composer}, or domain-specific
languages such as Amazon States Language~\cite{knix, aws_step_function}.
However, this approach has several limitations with regard to expressiveness, usability, and applicability.



\PHM{Limited expressiveness.}
Although the current function-oriented orchestration supports the workflows of
 simple invocation patterns, it
 becomes inconvenient or incapable of expressing more sophisticated
 function interactions, as summarized in Table~\ref{tab:expressive}.
 This is because the current function orchestration assumes that data flow in the 
 same way as how functions are invoked in a workflow, and that a function passes 
 its entire output to others by directly invoking them for immediate processing.
 These assumptions do not hold for many applications, hence developers resort to create workarounds.

\begin{figure}[]
	\centering
	\includegraphics[width=0.45\textwidth]{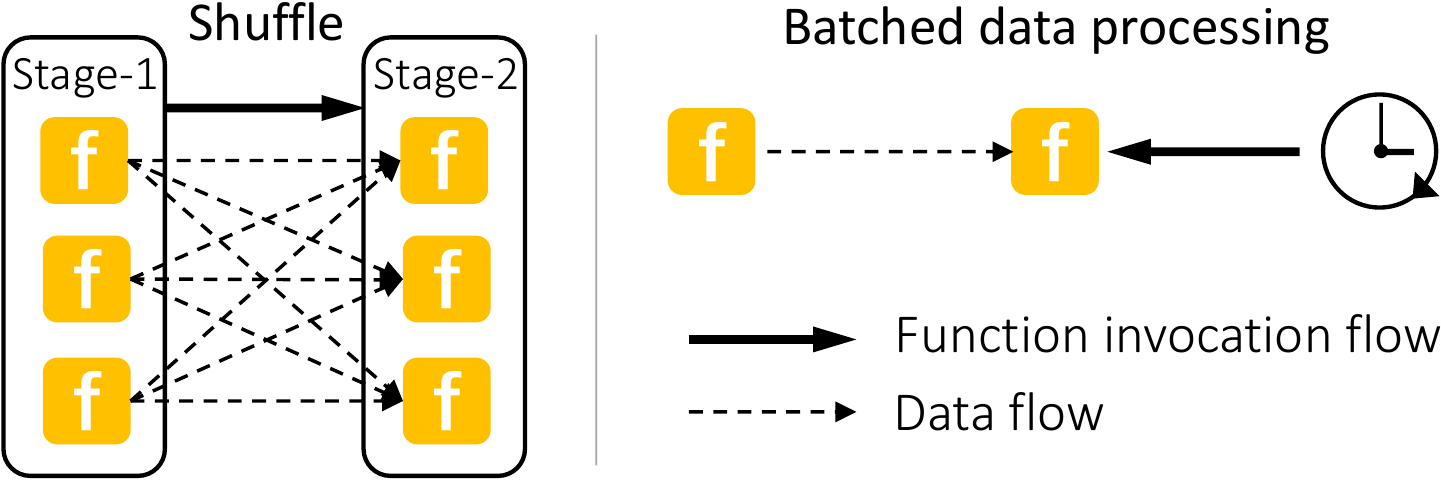}
	\caption{The shuffle operation (left) in data analytics and the batched data processing in a stream (right). }
	\label{fig:data_flow_example}
	\vspace{-.2in}
\end{figure}



For example, the ``shuffle'' operation in a data analytics job involves a
fine-grained, all-to-all data exchange between the functions of two stages
(e.g., ``map'' and ``reduce'' stages). As shown in Fig.~\ref
{fig:data_flow_example} (left), the output data of a function in stage-1
are shuffled and selectively redistributed to multiple functions in stage-2 based on
the output keys. However, the way to invoke functions is not the same as how the output data flow: only
after stage-1 completes can the workflow invoke all the stage-2 functions
in parallel. In current serverless platforms, developers must manually
implement such a complex data shuffle invocation via external storage~\cite
{klimovic_pocket:_2018,pu_shuffling_2019}, which is neither flexible nor
efficient.

Another example is a batched stream analytics job which periodically 
invokes a
function to process the data continuously received during a time window~\cite
{chintapalli_benchmarking_2016,xu_move_2021}, as shown in Fig.~\ref
{fig:data_flow_example} (right). A serverless workflow cannot effectively express
this invocation pattern as the function is not immediately triggered when the data
arrive, and thus developers have to rely on other cloud services (e.g., AWS
Kinesis~\cite{aws_kinesis}) to batch the data for periodic function
invocations~\cite{lambda_scenarios,lambda_stream,lambda_stream_example}. Note
that, even with the latest stateful workflow (e.g., Azure Durable
Functions~\cite{azure_durable_function_entity}), an addressable function needs
to keep running to receive data. As we will show in \S\ref
{sec:eval_app}, deploying a long-running function not only incurs extra
resource provisioning cost but results in an unsatisfactory performance.




\PHM{Limited usability.}
Current serverless platforms provide various options for 
data exchange between functions. Functions can exchange data either synchronously or
asynchronously via a message broker or a shared storage~\cite
{sreekanti_cloudburst_2020,aws_cache,aws_s3,knix,klimovic_pocket:_2018,
step_func_large_data}. They can also process data from various sources, such
as nested function calls, message queues, or other cloud services~\cite
{lambda_invoking}.

\begin{figure}[tbp]
	\centering
	\includegraphics[width=0.45\textwidth]{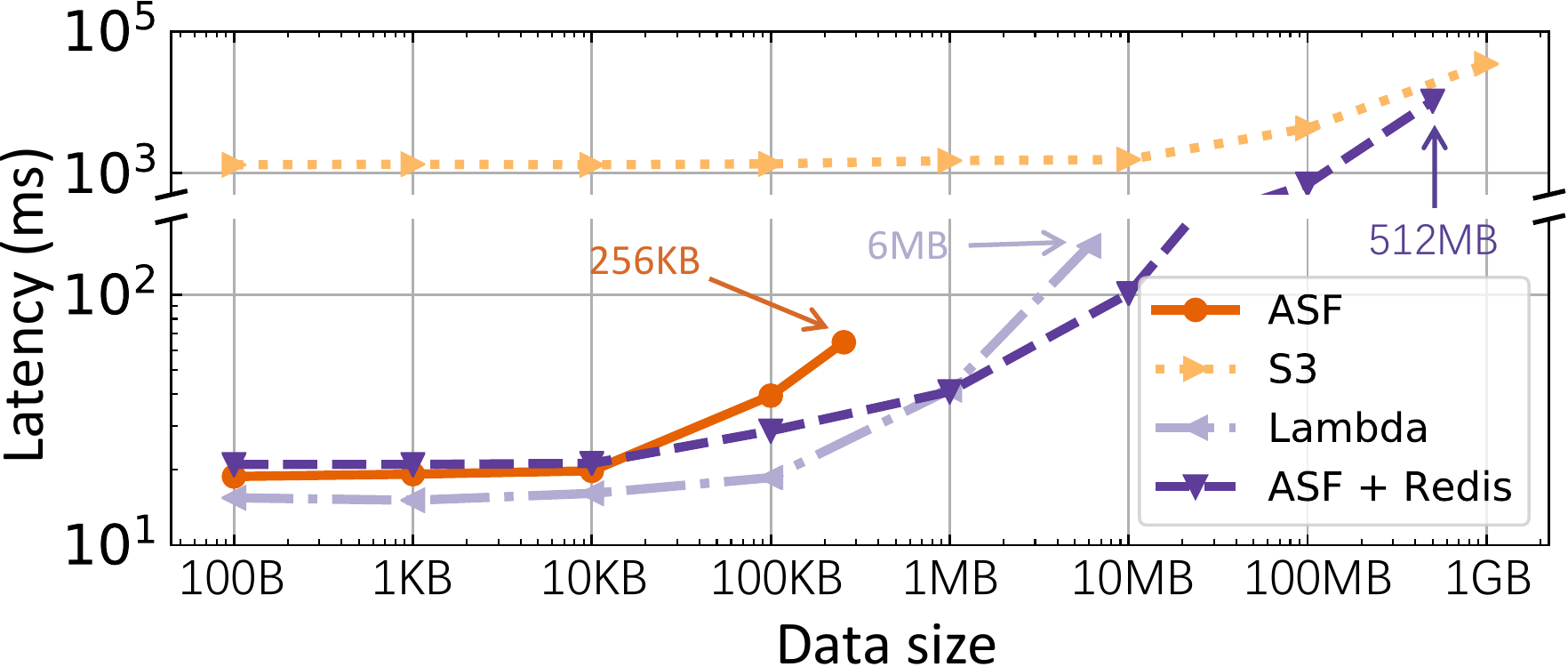}
	\caption{The interaction latency of two AWS Lambda functions under  various data sizes using four approaches.}
	\label{fig:aws_interact}
	\vspace{-.2in}
\end{figure}

The lack of a single best approach to exchange data between functions
significantly complicates the development
and deployment of serverless applications, as developers must find their own
ways to efficiently pass data across functions~\cite
{mahgoub_sonic_nodate} which can be dynamic and non-trivial; thus, reducing the usability of serverless platforms.
To illustrate this problem, we compare four data-passing approaches in
AWS Lambda: a) calling a function directly (Lambda), b) using
AWS Step Functions (ASF) to execute a two-function workflow\footnote{We use
the ASF Express Workflows in our experiments as it delivers higher performance
than the ASF Standard Workflows~\cite{step_func_express_workflow}.}, c) allowing
functions to access an in-memory Redis store for fast data
exchange (ASF+Redis), and d) configuring AWS S3 to invoke a function upon data
creation (S3)~\cite{lambda_invoke_with_s3}. Fig.~\ref
{fig:aws_interact} compares the latencies of these four approaches under
various data volumes.  Lambda is efficient for transferring small
data; ASF+Redis is efficient for transferring large data; 
the maximum data volume supported by each approach varies considerably,
and only the S3 approach can support virtually unlimited (but slow) data
exchange.  Thus, there is no single approach that prevails across all
scenarios, and developers must carefully profile the data patterns of their
applications and the serverless platforms to optimize the performance of data exchange between
interacting functions. 

To make matters worse, the data volume exchanged between functions
depends on the workload, which may be irregular or unpredictable. Thus, there
may be no best \emph{fixed} approach to exchanging data between interacting functions, and
developers have to write complex logic to select the best approach at
runtime. Developers also need to consider the interaction cost. Previous work
has highlighted the tricky trade-off between I/O performance and cost when
using different storage to share intermediate data~\cite
{pu_shuffling_2019,klimovic_pocket:_2018}, which further exacerbates the
usability issue. Altogether, these common practices bring a truly \emph
{non-serverless} experience to developers as they still have to deal with
server and platform characteristics.

\PHM{Limited applicability.}
Existing serverless applications are typically
not latency-sensitive.  This is because current serverless platforms usually
have a function interaction delay of multiple or tens of milliseconds (\S\ref
{sec:eval_func_interaction}), and such delays accumulate as more functions are 
chained together in an application workflow. For example, in AWS Step Functions, each
function interaction causes a delay of more than 20 ms, and the total
platform-incurred delay for a 6-function chain is over 100~ms, which may
not be acceptable in many latency-sensitive applications~\cite
{jia_nightcore_2021}. In addition, as current serverless platforms cannot
efficiently support the sharing of varying-sized data between functions
(as described earlier), they are ill-suited for data-intensive
applications~\cite
{aws_lambda,knix,jia_nightcore_2021,sreekanti_cloudburst_2020,
pu_shuffling_2019,fouladi_encoding_nodate}. Altogether, the above
characteristics substantially limit the applicability of current serverless
platforms.

\section{Data-Centric Function Orchestration}
\label{sec:orchestration}


In this section, we address the aforementioned limitations of the 
function orchestration practice in current serverless platforms, with a novel data-centric approach.
We will describe how this approach can be applied to develop a
new serverless platform later in \S\ref{sec:system}.

\subsection{Key Insight}
\label{sec:orch_idea}


As discussed in \S\ref{sec:deploy_app}, the current function orchestration
practice only specifies the high-level invocation dependencies between
functions, and thus has little fine-grained control over how these functions
exchange data. In particular, the current practice assumes the tight
coupling between function flows and data flows. Therefore, when a
function returns its result, the workflow has no knowledge about how
the result should be consumed (e.g., in full or part, directly or
conditionally, immediately or later). To address these limitations, 
\emph{an effective serverless platform must allow
fine-grained data exchange between the functions of a workflow, while
simultaneously providing a unified and efficient approach for function
invocation and data exchange}.


Following this insight, we propose a new \emph{data-centric approach} to function
orchestration. We note that intermediate data (i.e., results returned
by functions) are typically short-lived and immutable~\cite
{klimovic_pocket:_2018,tang_lambdata_2020}: after they are generated, they
wait to be consumed and then become obsolete.\footnote{For data that need
durability, they can be persisted to a durable storage.} We therefore make
data consumption explicit and enable it to trigger the target functions. Developers can
thus specify when and how intermediate data should be passed to the
target functions and trigger their activation, which can then drive the
execution of an entire workflow. As intermediate data are not updated once
they are generated~\cite{klimovic_pocket:_2018,tang_lambdata_2020}, using
them to trigger functions results in no consistency issues.

The data-centric function orchestration addresses the limitations
of the current practice via three key advances. First, it breaks the tight
coupling between function flows and data flows, as data do not have to follow
the exact order of function invocations. It also enables a flexible and fine-grained
control over data consumption, and therefore can express a rich set of workflow
patterns (i.e., \emph{rich expressiveness}). Second, the data-centric function orchestration
provides a unified programming interface for both function invocations and
data exchange, obviating the need for developers to implement complex logic via
a big mix of external services to optimize data exchange (i.e., \emph{high usability}).
Third, knowing when and how the intermediate data will be consumed provides
opportunities for the serverless platform scheduler to optimize the locality of functions and relevant data, 
and thus latency-sensitive and
data-intensive applications can be supported efficiently (i.e., \emph
{wide applicability}). 

\begin{figure}[]
	\centering
	\includegraphics[width=0.4\textwidth]{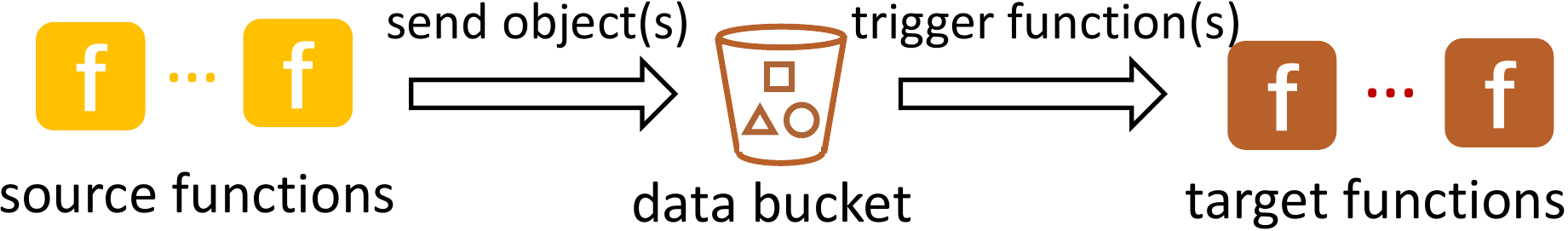}
	\caption{An overview of triggering functions in data-centric orchestration. Source functions send intermediate data to the associated bucket, which can be configured to automatically trigger target functions.}
	\label{fig:bucket_trigger}
\end{figure}

\subsection{Data Bucket and Trigger Primitives}
\label{sec:orch_design}

\PHB{Data bucket.} 
To facilitate the data-centric function orchestration, we design a \emph{data bucket}
abstraction and a list of \emph{trigger primitives}. Fig.~\ref
{fig:bucket_trigger} gives an overview of how functions are triggered. A serverless application creates one or multiple data buckets
that hold the intermediate data. Developers can configure each bucket with
triggers that specify when and how the data should invoke the target functions 
and be consumed by them. When executing a workflow, the source functions
directly send their results to the specified buckets. Each bucket checks if the
configured triggering condition is satisfied (e.g., the required data are
complete and ready to be consumed). If so, the bucket triggers the target functions
automatically and passes the required data to them. This
process takes place across all buckets, which collectively drive the
execution of an entire workflow.

We design various trigger primitives for buckets to specify how functions are
triggered. The interaction patterns between functions can be broadly
classified into three categories:

\vspace{.4em}
\noindent\textbf{Direct trigger primitive} (i.e., \texttt{Immediate}) allows one or
 more functions to directly consume data in the associated buckets.  This primitive has
 no specified condition, and triggers the target functions immediately once
 the data are ready to be consumed. This primitive can easily support sequential
 execution or invoke multiple functions in parallel (fan-out).


\vspace{.4em}
\noindent\textbf{Conditional trigger primitives}
trigger the target function(s) when the developer-specified conditions are met.
\begin{itemize}[topsep=0.5pt,itemsep=-.5ex]
\item \texttt{ByBatchSize}: It triggers the function(s) when the associated bucket has accumulated a certain number of data objects. It can be used to enable the batched stream processing~\cite{lambda_stream_example,lambda_stream} in a way similar to Spark Streaming.
\item \texttt{ByTime}: It sets up a timer and triggers the function(s) when the timer expires.
All the accumulated data objects are then passed to the function(s) as input. It can be used to implement routine tasks~\cite{xu_move_2021,chintapalli_benchmarking_2016}.
\item \texttt{ByName}: It triggers the function(s) when the bucket receives 
a data object of a specified name. It can be used to enable conditional 
invocations by choice~\cite{asf_choice}.
\item \texttt{BySet}: It triggers functions when a specified set of data objects are all complete and ready to be consumed. It can be used to enable the assembling invocation (fan-in).
\item \texttt{Redundant}: It specifies $n$ objects to be stored in a bucket and triggers
the function(s) when any $k$ of them are available and ready to be consumed. It can be used to execute redundant requests and perform late binding for straggler mitigation and improved reliability~\cite{vulimiri_more_2012,rashmi_ec-cache_2016,kosaian_parity_2019}.
\end{itemize}

\vspace{.4em}
\noindent\textbf{Dynamic trigger primitives}, 
unlike the previous two categories with statically-configured triggers, 
allow data exchange patterns to be configured at runtime.
\begin{itemize}[topsep=0.5pt,itemsep=-.5ex]
	\item \texttt{DynamicJoin}: It triggers the assembling functions when a set of data objects are ready, which can be dynamically configured at runtime. It enables the dynamic parallel execution like `\texttt{Map}' in AWS Step Functions~\cite{asf_map}.
	\item \texttt{DynamicGroup}: It allows a bucket to divide its data objects into multiple groups, each of which can be consumed by a set of functions. The data grouping is dynamically performed based on the objects' metadata (e.g., the name of an object). Once a group of data objects are ready, they trigger the associated set of functions.	
\end{itemize}


\begin{table}[tb]
	\centering
	\footnotesize
	\caption{Expressiveness comparison between the function-oriented workflow primitives in AWS Step Functions (ASF) and the data-centric trigger primitives in \SysName.}
\begin{tabular}{ m{10em} | m{6em} | m{6em}}
	\hline
	\textbf{Invocation Patterns} & \textbf{ASF} & \textbf{\SysName} \\
	\hline\hline
	Sequential Execution & \texttt{Task} & \texttt{Immediate} \\ 
	\hline
	Conditional Invocation & \texttt{Choice} & \texttt{ByName} \\ 
	\hline
	Assembling Invocation & \texttt{Parallel} & \texttt{BySet} \\ 
	\hline
	Dynamic Parallel & \texttt{Map} & \texttt{DynamicJoin} \\ 
	\hline
	Batched Data Processing & - & \texttt{ByBatchSize} \texttt{ByTime} \\ 
	\hline
	$k$-out-of-$n$ & - & \texttt{Redundant} \\ 
	\hline
	MapReduce & - & \texttt{DynamicGroup} \\ 
	\hline
\end{tabular}
\label{tab:expressive}
\end{table}

\begin{figure}[tbp]
    \centering
    \includegraphics[width=0.45\textwidth]{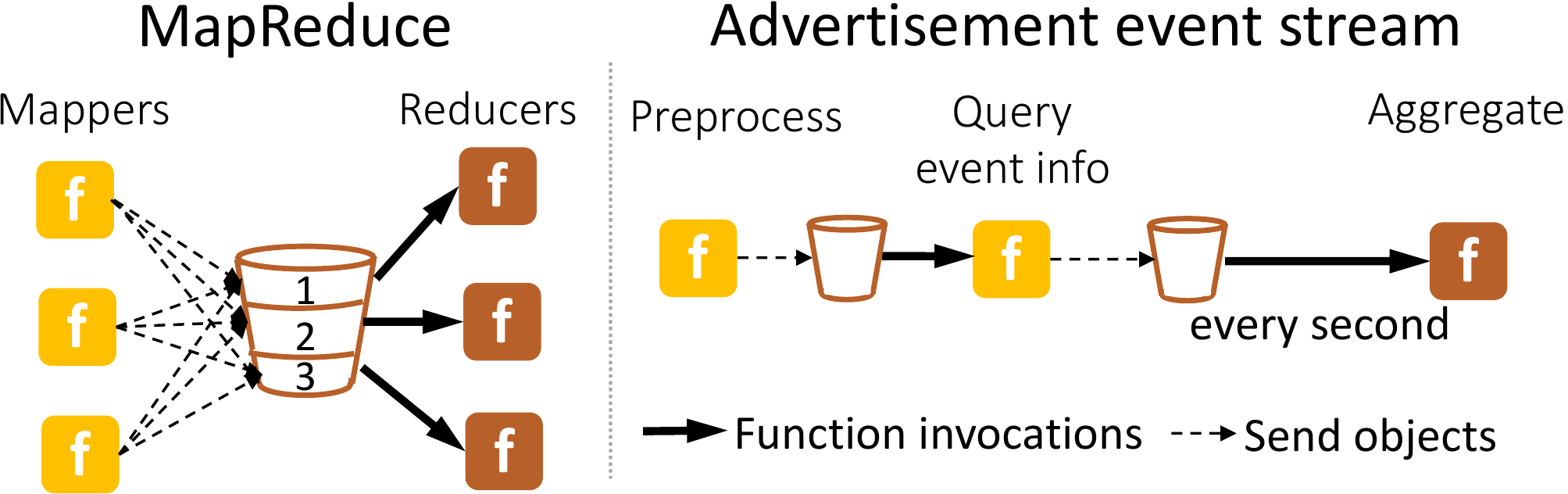}
    \caption{Usage examples of two primitives: \texttt{DynamicGroup} for data shuffling in MapReduce (left), and \texttt{ByTime} for periodic data aggregation in the event stream processing (right).}
    \label{fig:pri_all}
    \vspace{-.1in}
\end{figure}

Dynamic trigger primitives are critical to implement some widely-used computing frameworks, e.g., MapReduce
(which is hard to support in current serverless platforms as it requires
triggering parallel functions at every stage and optimizing the fine-grained,
all-to-all data exchange between them~\cite
{jonas_occupy_2017,pu_shuffling_2019,klimovic_pocket:_2018}, see \S\ref
{sec:deploy_app}). Here, our \texttt{DynamicGroup} primitive provides an easy solution to these
issues.  As shown in Fig.~\ref{fig:pri_all} (left), when a map function sends
intermediate data objects to the associated bucket, it also specifies to which
data group each object belongs (i.e., by specifying their associated
keys). Once the map functions are all completed, the bucket
triggers the reduce functions, each consuming a group of objects.


\begin{figure}[t!]
\begin{lstlisting}[style=trigger, escapechar=|]
struct BucketKey {
	string bucket_; // bucket name
	string key_; // key name
	string session_; // unique session id per request
};

abstract class Trigger {
	// Check whether to trigger functions for a new object.
	vector<TriggerAction> action_for_new_object(
			BucketKey bucket_key);
	
	// Notify the information of a source function.
	void notify_source_func(string function_name,
			string session, vector<string> function_args);
	
	// Check whether to re-execute source functions.
	vector<TriggerAction> action_for_rerun(string session);
};
\end{lstlisting}
\vspace{-.4in}
\caption{Three main methods of the trigger interface.}
\label{fig:script_trigger_interface}
\vspace{-.1in}
\end{figure}

\final{We have developed a new serverless platform, \SysName, which implements the aforementioned data bucket abstraction and trigger primitives.  The design of \SysName will be detailed in \S\ref{sec:system}. Table~\ref{tab:expressive} lists all the supported trigger primitives in current \SysName platform. Compared to AWS Step Functions (ASF), \SysName supports more sophisticated invocation patterns and provides richer
 expressiveness for complex workflows.
 We note that Azure Durable Functions~\cite{azure_durable_function} can also achieve rich expressiveness for complex workflows (\S\ref{sec:setup}). Yet, it fails to achieve the other two desired properties, i.e., high usability and wide applicability (\S\ref {sec:eval_app}).
 }

\PHM{\final{Abstract interface.}}
\final{\SysName's trigger primitives are not only limited to those
 listed in Table~\ref{tab:expressive}.
 Specifically, we provide an abstract interface for developers to implement 
 customized trigger primitives for their applications, if needed. 
Fig.~\ref{fig:script_trigger_interface} shows the three
 main methods of the trigger interface. The first method, \texttt
 {action\_for\_new\_object}, is provided to specify how
 the trigger's associated target functions should be invoked. This method can be called when a new
 data object arrives: it checks the current data status and returns a list of
 functions to invoke, if any. The method can also be called periodically in a
 configurable time period through periodical checking (e.g., \texttt
 {ByTime} primitive). The other two methods, \texttt
 {notify\_source\_func} and \texttt{action\_for\_rerun}, are provided to implement 
 the fault handling logic which re-executes the trigger's associated source functions in case of failures. 
 In particular, through \texttt{notify\_source\_func}, a trigger can obtain the information of a source function once the function starts, including the function name, session, and arguments;
 \SysName also performs the periodic re-execution checks by calling \texttt{action\_for\_rerun}, which returns a list of timeout functions, such that \SysName can then re-execute them.
 The detailed fault tolerance mechanism will be described in \S\ref{sec:system_fault_tolerance}.
\ifthenelse{\boolean{exclude_appendix}}{We give an example of implementing a customized  \texttt{ByBatchSize} trigger primitive via the abstract interface in our technical report~\cite{pheromone_full_version}.}{
Appendix~\ref{sec:appendix_trigger_interface} gives an example of implementing the \texttt{ByBatchSize} trigger primitive via the interface.}}

\subsection{Programming Interface}
\label{sec:orch_api}

\begin{table*}[tb]
	\centering
	\footnotesize
	\caption{The APIs of user library which developers  use to operate on intermediate data objects and drive the workflow execution.}
	
	\begin{tabular}{ m{7em} | m{22em} | m{28em}}
		\hline
		\textbf{Class} & \textbf{API} & \textbf{Description} \\
		\hline\hline
		\multirow{ 2}{*}{\texttt{EpheObject}} & \texttt{void* \textbf{get\_value}()} & Get a pointer to the value of an object. \\
		\cline{2-3}
		& \texttt{void \textbf{set\_value}(value, size)} & Set the value of an object.  \\ 
		\hline
		\multirow{ 5}{*}{\texttt{UserLibrary}} & \texttt{EpheObject* \textbf{create\_object}(bucket, key)} & Create an object by specifying its bucket and key name. \\
		& \texttt{EpheObject* \textbf{create\_object}(function)} & Create an object by specifying its target function. \\ 
		& \texttt{EpheObject* \textbf{create\_object}()} & Create an object. \\ 
		\cline{2-3}
		& \texttt{void \textbf{send\_object}(object, output=false)} & Send an object to its bucket, and set the \texttt{output} flag if it needs to persist. \\ 
		\cline{2-3}
		& \texttt{EpheObject* \textbf{get\_object}(bucket, key)} & Get an object by specifying its bucket and key name. \\ 
		\hline
	\end{tabular}
	\label{tab:api}
	\vspace{-.1in}
\end{table*}

Our \SysName serverless platform currently accepts functions written in C++, with capabilities to support more languages (see \S\ref{sec:discussion}).  \SysName also provides 
a Python client through which developers can program function interactions.

\PHM{Function interface.}
Following the common practice, developers implement their functions
through the \texttt{handle()} interface (see Fig.~\ref{fig:interface}), which
is similar to the C++ main function except that it takes a user library as
the first argument. The user library provides a set of APIs (see
Table~\ref{tab:api}) that allow developers to operate on intermediate data
objects. These APIs enable developers to create intermediate data objects
(\texttt{EpheObject}), set their values, and send them to the buckets. A data object
can also be persisted to a durable storage by setting the \texttt{output} flag when calling \texttt{send\_object()}. When a bucket receives
objects and decides to trigger next function(s), it automatically packages
relevant objects as the function arguments (see Fig.~\ref{fig:interface}). A
function can also access other objects via the \texttt{get\_object()} API.

\begin{figure}[tb]
\begin{lstlisting}[style=interface]
int handle(UserLibraryInterface* library,\
  int arg_size, char** arg_values);
\end{lstlisting}
\vspace{-.2in}
\caption{Function interface.}
\label{fig:interface}
\vspace{-.1in}
\end{figure}

\begin{figure}[tb]
\begin{lstlisting}[style=mypython, escapechar=|]
app_name = 'event-stream-processing'
bucket_name = 'by_time_bucket'
trigger_name = 'by_time_trigger'
prim_meta = {'function':'aggregate', 'time_window':1000}
re_exec_rules = ([('query_event_info', EVERY_OBJ)], 100)
client.create_bucket(app_name, bucket_name)
client.add_trigger(app_name, bucket_name, trigger_name, \
		BY_TIME, prim_meta, hints=re_exec_rules)
\end{lstlisting}
\vspace{-.4in}
\caption{Configuring a bucket trigger to periodically invoke a function in a stream processing workflow.}
\label{fig:script_configure_triggers}
\vspace{-.2in}
\end{figure}

\PHM{Bucket trigger configuration.}
Developers specify how the intermediate data should trigger functions in a
workflow via our Python client. The client creates buckets and configures
triggers on the buckets using the primitives described in~\S\ref
{sec:orch_design}. Functions can then interact with the buckets by creating,
sending and getting objects using the APIs listed in Table~\ref{tab:api}.

\new{
As an example, we refer to a stream processing workflow~\cite{chintapalli_benchmarking_2016}
as shown in Fig.~\ref{fig:pri_all} (right). This workflow
first filters the incoming advertisement
events (i.e., \texttt{preprocess}) and checks which campaign each event
belongs to (i.e., \texttt{query\_event\_info}).  It then stores the returned results into
a bucket and periodically invokes a function (i.e., \texttt{aggregate}) to count the
events per campaign every second. Fig.~\ref
{fig:script_configure_triggers} gives a code snippet of configuring a bucket
trigger that periodically invokes the \texttt{aggregate} function, where
a \texttt{ByTime} trigger is created with the primitive metadata that 
specifies both the target function and the triggering time
window (line 4). Developers can optionally specify
a re-execution rule in case of function failures, e.g., by re-executing 
the \texttt{query\_event\_info} function if the bucket has not received
\texttt{query\_event\_info}'s output in 100~ms (line 5).
We will describe the  fault tolerance and re-execution
in \S\ref{sec:system_fault_tolerance}.
\ifthenelse{\boolean{exclude_appendix}}{A full script of deploying this workflow is given in our technical report~\cite{pheromone_full_version}.}{Appendix~\ref{sec:appendix_aes} gives a full script of deploying this workflow.}}

To summarize, our data bucket abstraction, trigger primitives, and programming interface facilitate the data-centric function orchestration, and enable
developers to conveniently implement their application workflows and express various
types of data patterns and function invocations.  In addition, the unified programming
interface also obviates the need to make an ad-hoc selection from many APIs
provided by various external services, such as a message broker, in-memory
database, and persistent storage.

\section{\SysName System Design}
\label{sec:system}

This section presents the design of \SysName{}, a new serverless platform that
supports data-centric function orchestration. 


\begin{figure}[tbp]
    \centering
    \includegraphics[width=0.42\textwidth]{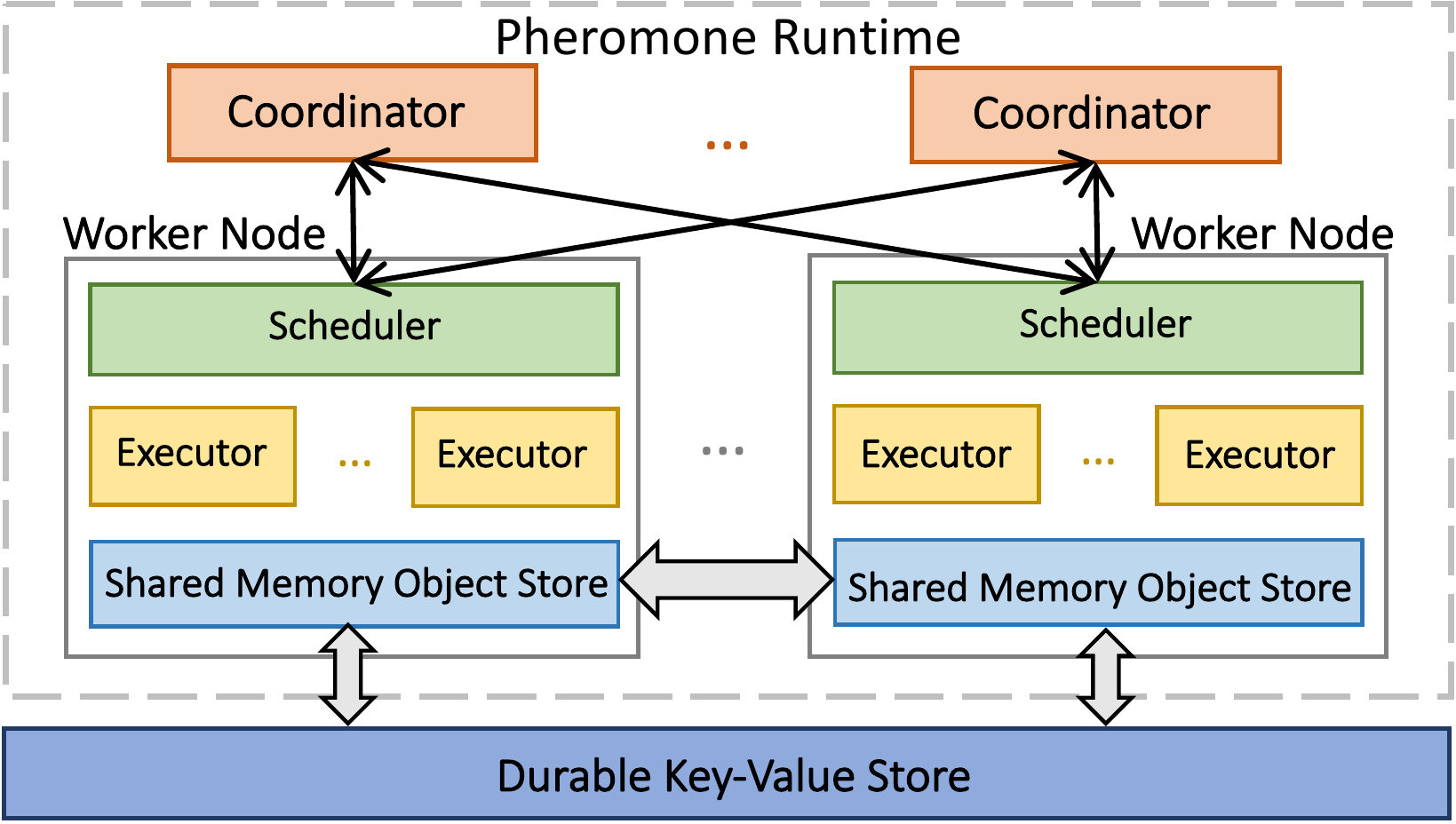}
    \caption{An architecture overview of \SysName{}.}
    \label{fig:architecture}
    \vspace{-.1in}
\end{figure}

\subsection{Architecture Overview}
\label{sec:sys_overview}

\revise{\SysName{} runs on a cluster of machines. Fig.~\ref{fig:architecture} shows an
 architecture overview. Each worker node follows instructions from a
 local scheduler, and runs multiple executors that load
 and execute the user function code as needed. A worker node also maintains a shared-memory object store
 that holds the intermediate data generated by functions. The object
 store provides a data bucket interface through which functions can
 efficiently exchange data within a node and with other nodes. It also
 synchronizes data that must persist with a remote durable key-value store,
 such as Anna~\cite{wu_anna_2018}. When new data are put into the
 object store, the local scheduler checks the associated bucket
 triggers. If the triggering conditions are satisfied, the local scheduler invokes the
 target function(s) either locally, or remotely with the help of
 a global coordinator that runs on a separate machine and performs cross-node coordination with a global view of bucket statuses.}


\subsection{Scalable Distributed Scheduling}
\label{sec:sys_schedule}

\revise{We design a two-tier, distributed scheduling scheme to exploit data locality and ensure high scalability, enabled by the data-centric approach.
Specifically, a workflow request first arrives at a global coordinator, which routes the request to a local scheduler on a worker node. 
The local scheduler invokes subsequent functions to locally execute the workflow whenever possible, thus reducing the invocation latency and incurring no network overhead.}


 \begin{figure}[tbp]
    \centering
    \includegraphics[width=0.46\textwidth]{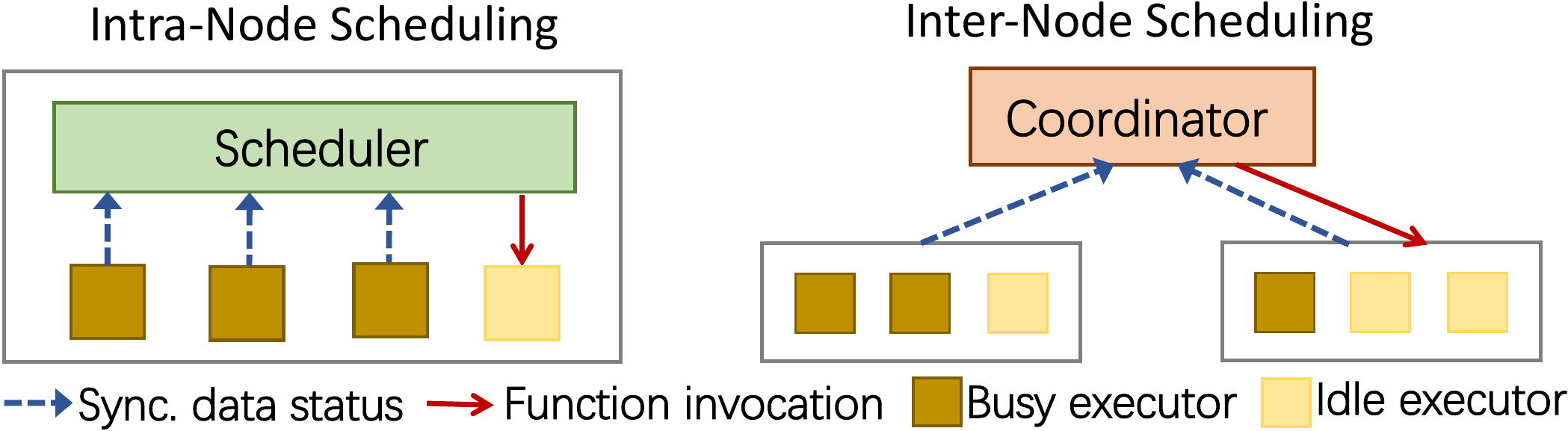}
    \caption{Intra-node (left) and inter-node (right) scheduling.}
    \label{fig:scheduling}
    \vspace{-.2in}
\end{figure}

\PHM{Intra-node scheduling.}
\revise{In \SysName{}, a local scheduler uses bucket triggers to invoke functions \emph{as
locally as possible}. The scheduler starts the first function of a
workflow and tracks its execution status via its bucket. The
downstream functions are triggered immediately on the same node when their expected data
objects are put into the associated buckets and ready to be consumed. As no cross-node communication is involved,
it reduces the function invocation latency and enables efficient consumption of data objects in a local workflow execution.
Fig~\ref
{fig:scheduling} (left) shows how the local scheduler interacts with executors when
running a workflow locally. The executors synchronize the data status
(e.g., the readiness of local objects in buckets) with the local scheduler, which
then checks the associated bucket triggers and invokes downstream functions
if the triggering conditions are met. The low-latency message exchange
between the scheduler and executors is enabled via an on-node shared-memory
object store.}

A local scheduler makes scheduling decisions based on the status of executors. The
scheduler only routes function requests to idle executors that have no
running tasks, avoiding concurrent invocations and resource
contention in each executor (similar to the concurrency model in AWS
Lambda~\cite{lambda_execution_model}). When the executor receives a request
for the first time, it loads the function code from the local object store
and persists it in memory for reuse in subsequent invocations. In case of
multiple idle executors, the scheduler prioritizes those with function code
already loaded to enable a warm start.\footnote{Many techniques have been
proposed to deal with cold starts of executors~\cite
{agache_firecracker_2020,du_catalyzer_2020,wang_faasnet_nodate,oakes_sock_2018,cadden_seuss_2020,shahrad_serverless_2020,fuerst_faascache_2021},
which can be applied directly in \SysName.}

\PHM{Delayed request forwarding from overloaded nodes.}
If the requests received by a local scheduler exceed the capacity of local
executors, the scheduler forwards them to a global coordinator, which
routes them to other worker nodes with sufficient resources. Instead of
forwarding the exceeding requests immediately, the scheduler waits for a
configurable short time period: if any local executors become
available during this period, the requested functions start and the requests
are served locally. The rationale is that it typically takes little time for executors to
become available as most serverless functions are short-lived~\cite
{shahrad_serverless_2020}, plus \SysName{} has microsecond-scale invocation
overhead (\S\ref{sec:eval_func_interaction}). Such a delayed scheduling
has proven effective for improving data locality~\cite{zaharia_delay_2010}.

\PHM{Inter-node scheduling.}
A global coordinator not only forwards requests from overloaded nodes to non-overloaded
nodes, but also drives the execution of a large workflow which needs to run across
multiple worker nodes that collectively host many functions of the workflow. This 
cannot be orchestrated by individual local schedulers without a global view.

\new{As Fig.~\ref{fig:scheduling} (right) shows, a coordinator gathers the
associated bucket statuses of the functions of a large workflow from multiple worker nodes, and triggers
the next functions as needed. Each node immediately synchronizes local bucket status with the coordinator upon any change, such that the coordinator maintains an up-to-date global view.
When the coordinator decides to trigger functions, it also updates this message to relevant workers, which reset local bucket status accordingly. 
This ensures a function invocation is neither missed nor duplicated.
Note that, some bucket triggers (e.g., \texttt{ByTime}) can only be performed at the coordinator with its global view; here, worker nodes only update their local statuses to the coordinator without checking trigger conditions.}

The data-centric orchestration
improves data locality in the inter-node scheduling. The coordinator makes
scheduling decisions using the node-level knowledge reported by local
schedulers, including cached functions, the number of idle executors, and the number of
objects relevant to the workflow. It then schedules a request to a
worker node with sufficient warm executors and the most relevant data objects.

\PHM{\new{Scaling distributed scheduling with sharded coordinators.}}
\new{\SysName employs a \emph{shared-nothing} model to significantly reduce
 synchronization between local schedulers and global coordinators, thus
 attaining high scalability. Specifically, it partitions the workflow
 orchestration tasks across \emph{sharded coordinators}, each of which
 manages a disjoint set of workflows. When executing a workflow, the
 responsible coordinator sends the relevant bucket triggers to a selected set
 of worker nodes and routes the invocation requests to them. A worker node
 may run functions of multiple workflows. For each workflow, 
 its data and trigger status are synchronized with the responsible
 coordinator only. This design substantially reduces communication and
 synchronization overheads, and can be achieved by running a standard cluster
 management service (e.g., ZooKeeper~\cite
 {zookeeper-paper,zookeeper-system}) that deals with coordinator failures and
 allows a client to locate the coordinator of a specific workflow. The client
 can then interact with this coordinator to configure data triggers and send
 requests. This process is automatically done by the provided client library
 and is transparent to developers.}

\subsection{Bucket Management and Data Sharing}
\label{sec:sys_data_sharing}

\new{We next describe how \SysName manages data objects in buckets,
and enables fast data sharing between functions.}

\PHM{\new{Bucket management.}}
\SysName{} uses an on-node shared-memory object store to maintain data
 objects, such
 that functions can directly access them via pointers (i.e., \texttt{EpheObject} in
 Table~\ref{tab:api}). A data object is marked ready when the source
 function puts it into a bucket via \texttt{send\_object()}. 
 The bucket can be distributed across its responsible coordinator and a number of worker nodes, where each worker node tracks local data status while the coordinator holds a global view (\S\ref{sec:sys_schedule}).
 Bucket status synchronization is only needed between the responsible coordinator and workers, as local statuses at different workers track their local objects only and are disjoint.

\SysName{} garbage-collects the intermediate objects of a workflow execution after 
 the associated invocation request has been \emph{fully} served along the workflow.
 In case a workflow is executed across
 multiple worker nodes, the responsible coordinator notifies the local scheduler on
 each node to remove the associated objects from its object store.
 \ifthenelse{\boolean{exclude_appendix}}{}{ Appendix~\ref{sec:eval_gc} gives an evaluation of the garbage collection mechanism.} 

 

\new{
When a worker node's local object store runs out of memory, a remote key-value store is
used to hold the newly generated data objects at the expense of an increased
data access delay.\footnote{Our current implementation does not support
spilling in-memory objects to disk, which we leave for future work.} Later,
when more memory space is made available (e.g., via garbage
collection), the node remaps the associated buckets to the local object store. 
In case a data object is lost due to system failures, \SysName automatically
re-executes the source function(s) to get it recovered (details in \S\ref
{sec:system_fault_tolerance}).}

\PHM{\new{Fast data sharing.}}
\revise{\SysName further adopts optimizations to fully reap the benefits of data locality enabled by its data-centric design.
As intermediate data are typically short-lived and immutable~\cite
 {klimovic_pocket:_2018,tang_lambdata_2020}, we trade their durability for
 fast data sharing and low resource footprint. With an on-node shared-memory
 object store, \SysName{} enables \emph{zero-copy} data
  sharing between local functions by passing only the pointers of 
 data objects to the target functions.
 This avoids the significant
 data copying and serialization overheads, and substantially reduces the latency of accessing local data objects.}

\new{To efficiently pass data to remote functions, \SysName{} also enables the
 \emph{direct} transfer of data objects between nodes. A function packages the
 metadata (e.g., locator) of a data object into a function request being sent to a remote node.
 The target function on the remote node uses such metadata to 
 directly retrieve the required data object. 
 Compared with using a remote
 storage for cross-node data sharing, our direct data transfer avoids unnecessary data
 copying, and thus leads to reduced network and storage overheads. While the
 remote-storage approach can ensure better data durability and consistency~\cite
 {knix,sreekanti_cloudburst_2020,shillaker_faasm_2020}, there is no such need
 for intermediate data objects. Only when data are specified to persist
 will \SysName{} synchronize data objects with a durable key-value store
 (see \texttt{send\_object()} in Table~\ref{tab:api}).}

\revise{
Note that, \SysName's data-centric design can expose details of intermediate data (e.g., the size of each data object), therefore we can further optimize cross-node data sharing.
For large data objects, they are sent as raw
byte arrays to avoid serialization-related overheads, thus significantly improving the performance of transferring large objects (see Fig.~\ref{fig:breakdown} in \S\ref{sec:eval_func_interaction}). 
For small data objects, Pheromone implements
a shortcut to transfer them between nodes: it
piggybacks small objects on the function invocation requests
forwarded during the inter-node scheduling (see \S\ref{sec:sys_schedule}). This
shortcut saves one round trip as the target function does not
need to additionally retrieve data objects from the source function. In addition, Pheromone runs an I/O thread pool on each
worker node to improve cross-node data sharing performance.
}





\subsection{Fault Tolerance}
\label{sec:system_fault_tolerance}

\revise{\SysName sustains various types of system component failures. In case an executor
 fails or a data object is lost, \SysName restarts the failed
 function to reproduce the lost data and resume the interrupted workflow.
 This is enabled by using the data bucket to re-execute its source function(s) if the expected output has not been received in a configurable timeout.
 This fault handling approach is a natural fit for data-centric function orchestration and brings two benefits.
 First, it can simplify the scheduling logic as data buckets can autonomously track the runtime status of each function and issue re-execution requests whenever necessary, without needing schedulers to handle function failures.
 Second, it allows developers to customize function re-execution rules when configuring data buckets, e.g., timeout.
 Fig.~\ref{fig:script_configure_triggers} gives an example of specifying re-execution rules (line 5).
 Fig.~\ref{fig:script_trigger_interface} shows the interface to implement the logic of function re-execution for a bucket trigger (\texttt
 {notify\_source\_func} and \texttt{action\_for\_rerun}).


}


\new{\SysName also checkpoints the scheduler state (e.g., the
 workflow status) to the local object store, so that it can quickly recover
 from a scheduler failure on a worker node. In case that an entire
 worker node crashes, \SysName re-executes the failed workflows on
 other worker nodes. \SysName can also handle failed coordinators with a standard
 cluster management service, such as ZooKeeper, as explained in \S\ref
 {sec:sys_schedule}.}



\begin{figure*}[tbp]
	\centering
	\includegraphics[width=.98\textwidth]{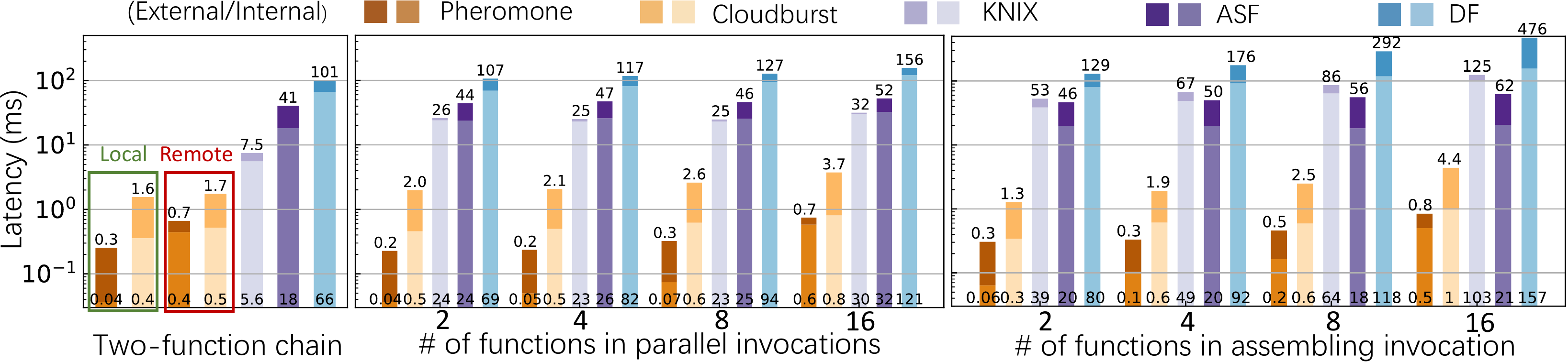}
	\caption{Latencies of invoking no-op functions under three interaction patterns: function chain, parallel and assembling invocations. Each bar is broken into two parts which measure the latencies of external (darker) and internal (lighter) invocations, respectively. 
		The overall latency value is given at the top of the bar, and the internal invocation latency
		is given at the bottom.}
	\label{fig:microbench_1}
	\vspace{-.1in}
\end{figure*}

\section{Implementation}
\label{sec:implementation}

\new{We have implemented \SysName atop Cloudburst~\cite{sreekanti_cloudburst_2020}, a lightweight, performant serverless platform. 
We heavily re-architected Cloudburst and implemented \SysName's key components (Fig.~\ref{fig:architecture}) in 5k lines of C++ code. 
These components were packaged into Docker~\cite{docker} images for  ease of deployment.  
\SysName's client was implemented in 400 lines of Python code.
Like Cloudburst, \SysName runs in a Kubernetes~\cite{kubernetes} cluster for convenient container management, and uses Anna~\cite{wu_anna_2018,wu_autoscaling_2019}, an autoscaling key-value store, as the durable key-value storage.
On each worker node, we mount a shared in-memory volume between containers for fast data exchange and message passing.
The executor loads function code as dynamically linked libraries, which is pre-compiled by developers and uploaded to \SysName{}. The entire codebase of \SysName is
open-sourced at~\cite{Pheromone-code}.}

\section{Evaluation}
\label{sec:evaluation}

In this section, we evaluate \SysName{} via a cluster deployment in AWS EC2.
Our evaluation answers three questions:
\begin{itemize}[topsep=0.5pt,itemsep=-.5ex]
    \item How does \SysName improve function interactions (\S\ref {sec:eval_func_interaction}) and ensure high scalability (\S\ref{sec:eval_scalability})?
    \item Can \SysName effectively handle failures (\S\ref{sec:eval_ft_gc})?
    \item Can developers easily implement real-world applications with \SysName{} and deliver high performance (\S\ref{sec:eval_app})?
\end{itemize}

\subsection{Experimental Setup}
\label{sec:setup}

\PHB{Cluster settings.}
We deploy \SysName{} in an EC2 cluster. The coordinators
run on the \texttt{c5.xlarge} instances, each with 4 vCPUs and 8~GB memory. Each
worker node is a \texttt{c5.4xlarge} instance with 16 vCPUs and 32~GB
memory. The number of executors on a worker node is configurable and we tune
it based on the requirements of our experiments. We deploy up to 8
coordinators and 51 worker nodes, and run clients on separate instances in
the same \texttt{us-east-1a} EC2 zone.

\PHM{\new{Baselines.}}
\new{We compare \SysName{} with four baselines.}

\new{
\emph{1) Cloudburst:} As an open-source platform providing fast state sharing, Cloudburst~\cite{sreekanti_cloudburst_2020} adopts \emph{early binding} in scheduling: it schedules all functions of a workflow before serving a request, and enables direct communications between  functions. It also uses function-collocated caches.
 As \SysName's cluster setting is similar to Cloudburst's, we deploy the two
 platforms using the same cluster configuration and resources.}

\new{
\emph{2) KNIX:} As an evolution of SAND~\cite{akkus_sand:_2018}, KNIX~\cite
 {knix} improves the function interaction performance by executing functions of a
 workflow as processes in the same container.  Message passing and data sharing can be done
 either via a local message bus or via a 
  remote persistent storage.}

\new{
\emph{3) AWS Step Functions (ASF):} We use ASF Express
 Workflows~\cite{step_func_express_workflow} to orchestrate function instances as it achieves faster function interactions than the ASF Standard Workflows~\cite{step_func_express_workflow}. As ASF
 has a size limit of transferring intermediate data (see Fig.~\ref
 {fig:aws_interact}), we use Redis~\cite{aws_cache}, a fast in-memory storage service, to share 
 large data objects between functions.}

\new{
\emph{4) Azure Durable Functions (DF):} Compared with ASF, DF provides a more
 flexible support for function interactions. It allows developers to express workflows in program code and offers the Entity Functions~\cite{azure_durable_function_entity} that can manage workflow states following the actor model~\cite
 {bykov_orleans_2011,moritz_ray_2018}. 
 We include DF to study whether this expressive orchestration approach can achieve satisfactory performance.}

\new{Here, Cloudburst, KNIX and ASF focus more on optimizing function interactions of a workflow, while DF provides rich expressiveness.
Note that, for the two commercial platforms, i.e., ASF and DF, we cannot control their orchestration runtime.
To make a fair comparison, we configure their respective Lambda and Azure functions such that the number of function instances matches that of executors in \SysName{}. The resource allocations of each function instance and
executor are also maintained the same. In our experiments, functions are all 
warmed up to avoid cold starts in all platforms.}




\subsection{Function Interaction}
\label{sec:eval_func_interaction}

\PHB{Function invocation under various patterns.}
We first evaluate the overhead of invoking no-op functions without any payload. We consider three common invocation patterns: sequential
execution (e.g., a two-function chain), parallel invocation (fan-out), and
assembling invocation (fan-in). We vary the number of involved functions for
parallel and assembling invocations to control the degree of parallelism.
Fig.~\ref{fig:microbench_1} shows the latencies of invoking no-op functions
under these three patterns.  Each latency bar is further broken down into the overheads of external and
internal invocations. The former measures the latency between the arrival of a
request and the complete start of the workflow, and the latter measures
the latency of internally triggering the downstream function(s) following the
designated pattern. In \SysName{}, the external invocation latency is mostly
due to the overhead of request routing which takes about 200
$\mu$s~\cite{aws_network_perf}. Note that, functions can be invoked locally or remotely
in \SysName{} and Cloudburst, thus we measure them respectively in Fig.~\ref
{fig:microbench_1}.
In our experiments, we report the average latency over 10 runs.


\begin{figure}
	\centering
	\includegraphics[width=.45\textwidth]{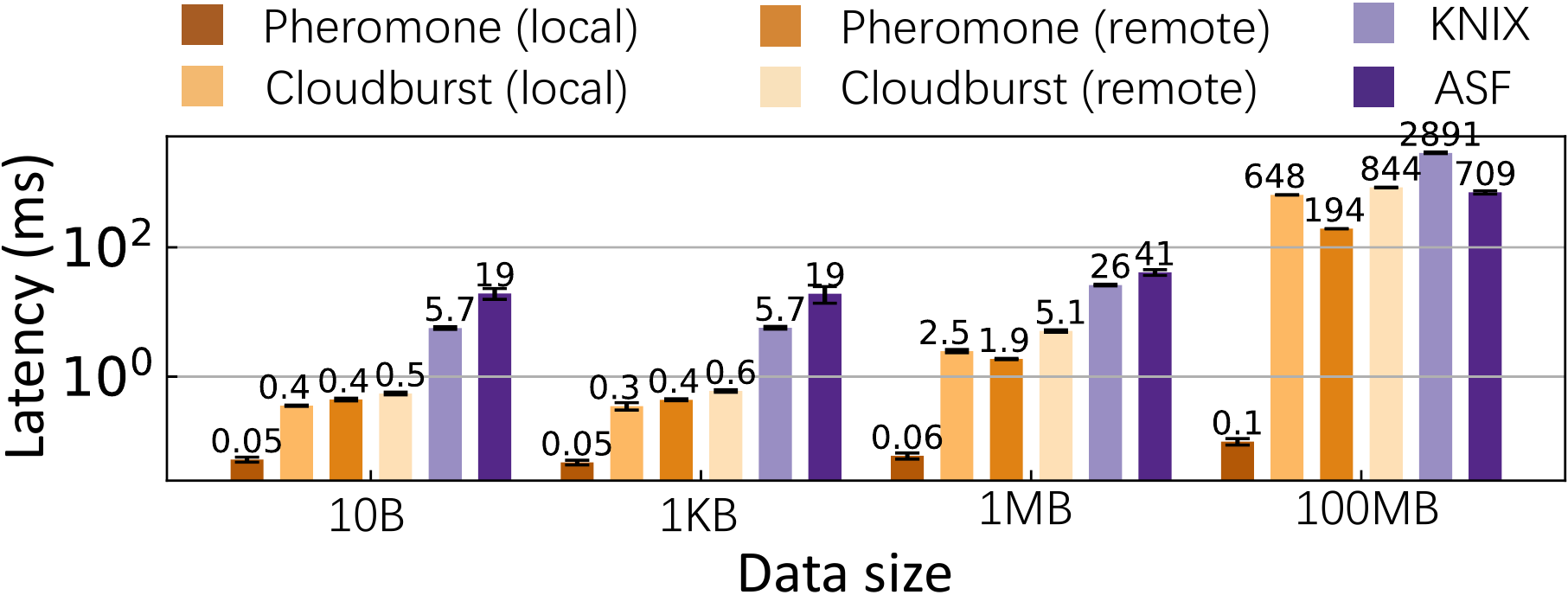}
	\caption{Latencies of a two-function chain invocation under various data sizes.}
	\label{fig:microbench_2}
	\vspace{-.1in}
\end{figure}

Fig.~\ref{fig:microbench_1} (left) compares the invocation latencies of a
 two-function chain measured on five platforms.  \SysName{} substantially
 outperforms the others. In particular, \SysName's shared memory-based message
 passing (\S\ref{sec:sys_data_sharing}) only incurs an overhead of less than 20 $\mu$s, reducing the local invocation latency to 40 $\mu$s, which is 10$\times$ faster than
 Cloudburst.
 The latency improvements become significantly more salient
 compared with other platforms (e.g., 140$\times$ over KNIX, 450$\times$
 over ASF). When invoking a remote function,
 both \SysName{} and Cloudburst require network transfer, leading to a
 similar internal invocation latency. Yet, Cloudburst incurs higher overhead
 than \SysName{} for external invocations as it needs to schedule the entire
 workflow's functions before serving a request (early binding), thus
 resulting in worse overall performance.

Fig.~\ref{fig:microbench_1} (center) and (right) show the invocation latencies under parallel and assembling invocations, respectively.
We also evaluate the cross-node function invocations in \SysName{} and Cloudburst by configuring 12 executors on each worker, thus forcing remote invocations when running 16 functions.
\SysName{} constantly achieves the best performance and incurs only sub-millisecond latencies in all cases, even for cross-node function invocations.
In contrast, Cloudburst's early-binding design incurs a much longer latency for function invocations as
the number of functions increases.
Both KNIX and ASF incur high invocation overheads in the parallel and assembling
scenarios. 
DF yields the worst performance, and we exclude it from the experiments hereafter.

\begin{figure}
    \centering
	\includegraphics[width=.45\textwidth]{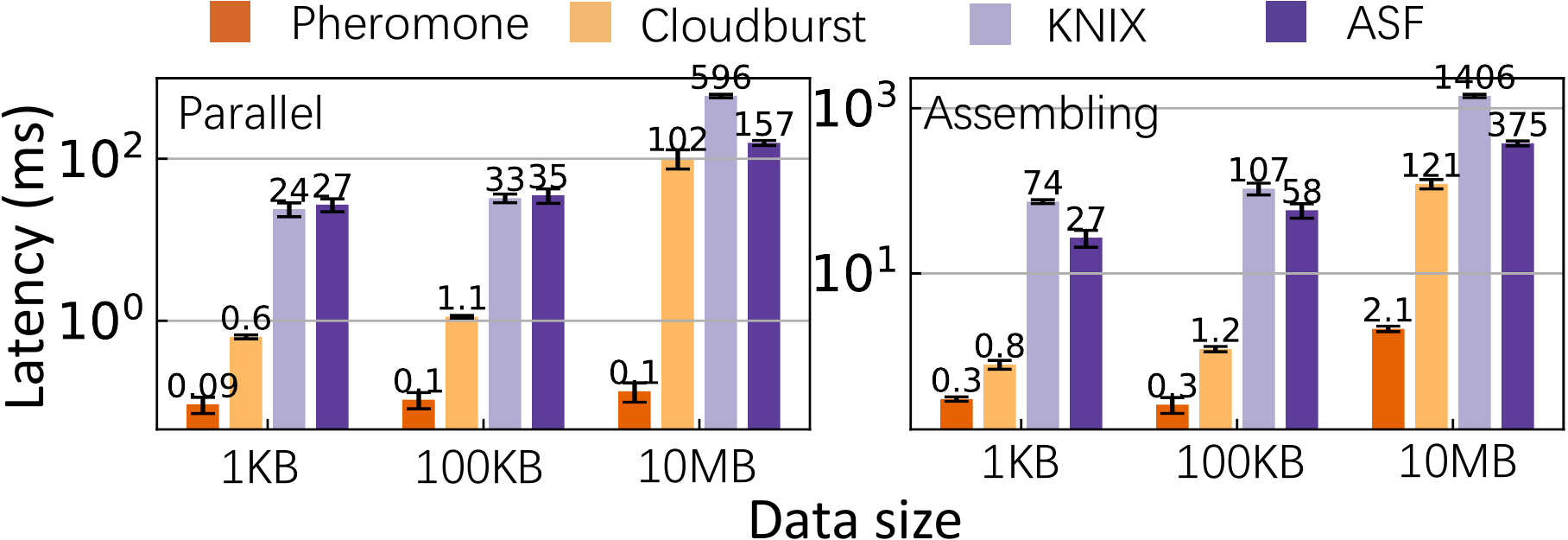}
	\caption{Latencies of parallel (left) and assembling (right) invocations under various data sizes, using 8 functions.}
	\label{fig:microbench_3}
	\vspace{-.1in}
\end{figure}

\PHM{Data transfer.}
We next evaluate the interaction overhead when transferring data between
functions. Fig.~\ref{fig:microbench_2} shows the invocation latencies of a
two-function chain with various data sizes in \SysName
{}, Cloudburst, KNIX, and ASF. We evaluate both local and remote data transfer
for \SysName{} and Cloudburst. For KNIX and ASF where the data transfer can be
done via either a workflow or a shared storage (i.e., Riak and Redis),
we report the best of the two choices.

For local data transfer, \SysName
{} enables zero-copy data sharing, leading to extremely low overheads
regardless of the data size, e.g., 0.1~ms for 100~MB data. In comparison,
Cloudburst needs the data copying and serialization, causing much longer latencies especially for large data objects.
For remote data transfer, both \SysName{} and Cloudburst support direct data
sharing across worker nodes. \SysName{} employs an optimized
implementation without (de)serialization, making it more efficient than Cloudburst.
Collectively, compared with \SysName, the serialization overhead in Cloudburst dominants the latencies of both local and remote invocations under large data exchanges, which diminishes the performance benefits of data locality: saving the cost of transferring 100~MB data across network only reduces the latency from 844~ms to 648~ms.
Fig.~\ref{fig:microbench_2} also shows that KNIX and ASF incur much longer latencies. While KNIX outperforms ASF when data objects are small, ASF becomes more efficient for passing large data because it is configured in our experiments to use the fast Redis in-memory storage for large data transfer.

We further evaluate the overhead of data transfer under parallel and
assembling invocations. For parallel invocation, we measure the
latency of a function invoking parallel downstream functions and passing data
to all of them; for assembling invocation, we measure the latency between
the transfer of the first object and the reception of all objects in the
assembling function. Fig.~\ref{fig:microbench_3} shows the latencies of these
two invocation patterns under various data sizes. 
Similarly, \SysName{} constantly achieves faster data
transfer compared with all other platforms for both invocation patterns.

\begin{figure}
    \centering
    \includegraphics[width=.4\textwidth]{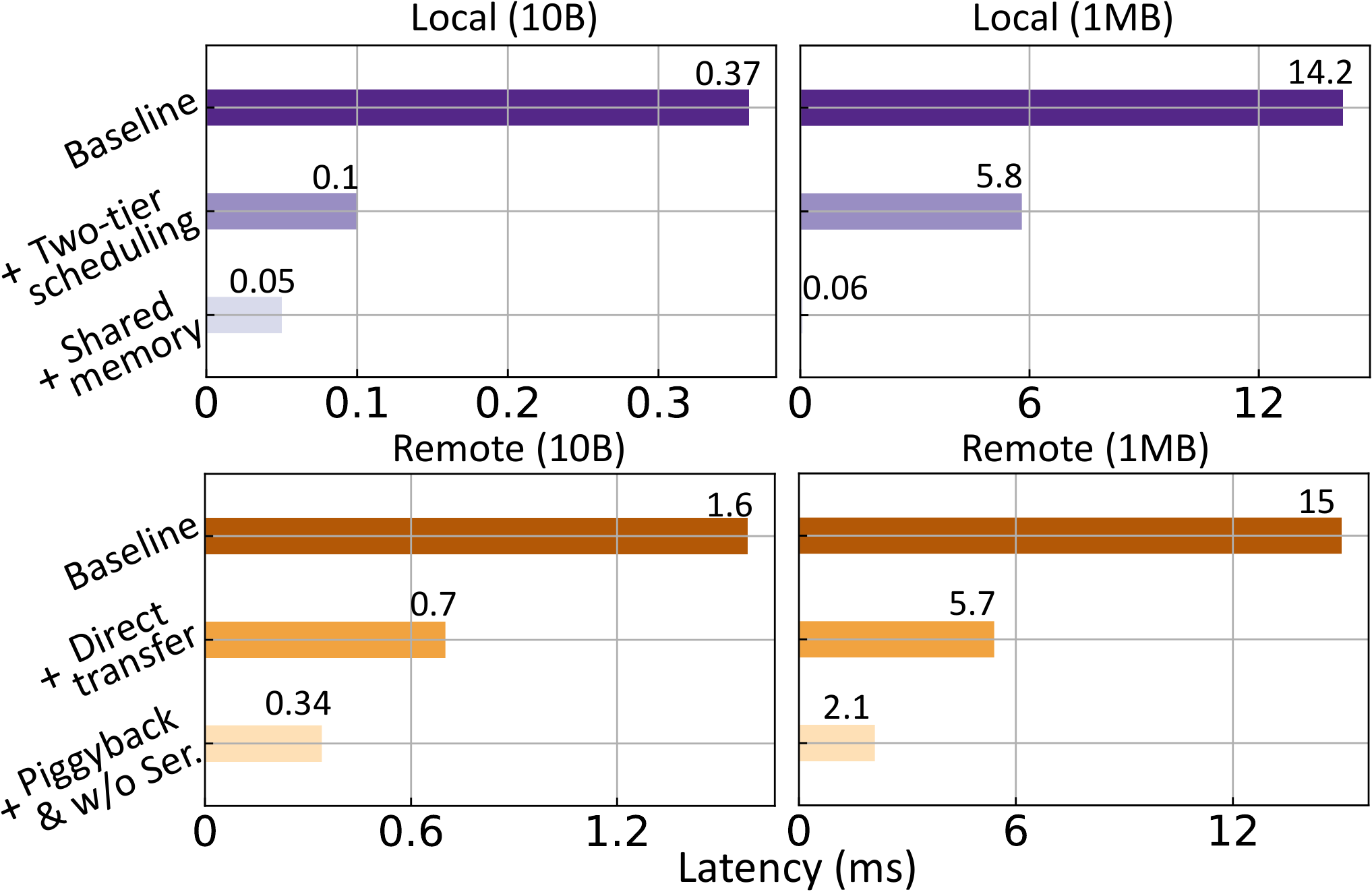}
    \caption{Improvement breakdown for local (top) and remote (bottom) invocations. Each case includes transferring 10~B (left) and 1~MB (right) of data in function invocations.}
    \label{fig:breakdown}
    \vspace{-.06in}
\end{figure}

\begin{figure}
	\centering
	\includegraphics[width=.4\textwidth]{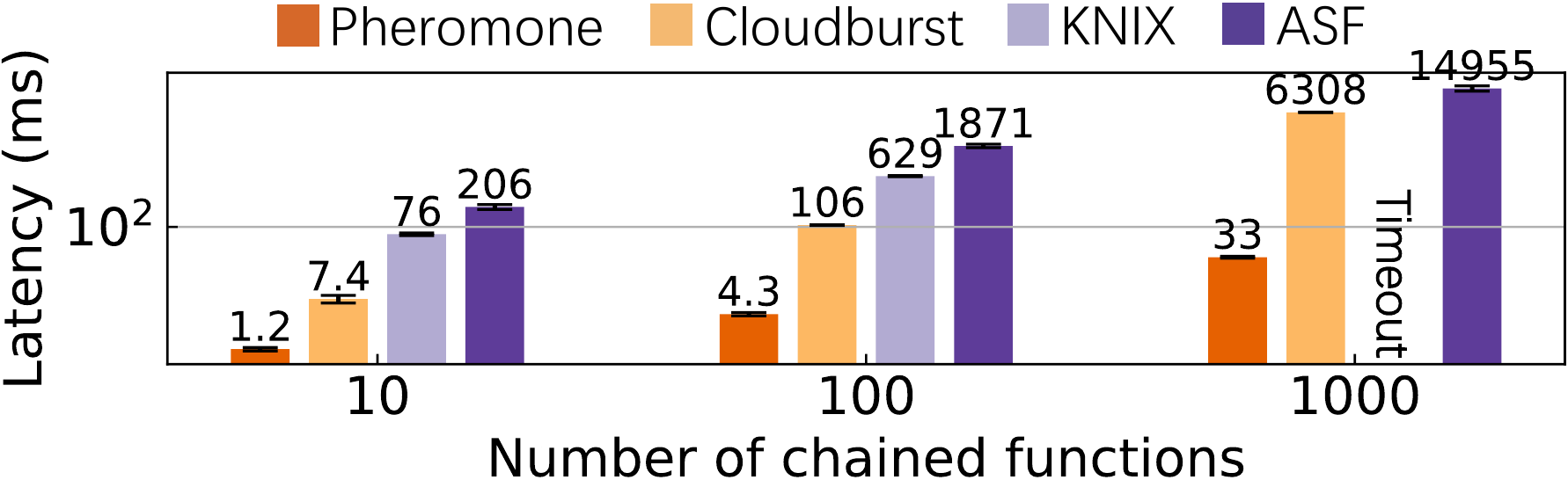}
	\caption{Latencies of function chains of different lengths.}
	\label{fig:function_chain}
	\vspace{-.1in}
\end{figure}

\PHM{\new{Improvement breakdown.}}
\new{To illustrate how each of our individual designs contributes to the performance improvement, we break down \SysName's function invocation performance and depict the
results in Fig.~\ref{fig:breakdown}.
Specifically, for local invocation, ``Baseline'' uses a central coordinator to invoke downstream functions (i.e., no local schedulers), which is today's common practice~\cite{aws_step_function};
``Two-tier scheduling'' uses our local schedulers for fast invocations on the same worker node (\S\ref{sec:sys_schedule}), where intermediate data objects are cached in the scheduler's memory and get copied to next functions;
``Shared memory'' further optimizes the performance via zero-copy data sharing (\S\ref{sec:sys_data_sharing}). 
Fig.~\ref{fig:breakdown} (top) shows that applying two-tier scheduling can reduce network round trips and achieve up to 3.7$\times$ latency improvement over ``Baseline''. Applying shared memory avoids data copy and serialization, further speeding up the data exchange especially for large objects (e.g., 1~MB) by two orders of magnitude.}

\new{For remote invocation, ``Baseline'' uses a durable key-value store (i.e., Anna~\cite{wu_anna_2018}) to exchange intermediate data among cross-node functions;
``Direct transfer'' reduces the communication overhead by allowing direct data passing between nodes (\S\ref{sec:sys_data_sharing}), where raw data objects on a node are serialized into a protobuf~\cite{protobuf} message and then sent to downstream functions;
``Piggyback \& w/o Ser.'' further optimizes the data exchange by piggybacking small objects on forwarded function requests and eliminating serialization (\S\ref{sec:sys_data_sharing}).
As shown in Fig.~\ref{fig:breakdown} (bottom), direct data transfer avoids interactions with the remote storage and improves the performance by up to 2.6$\times$ compared with baseline. The piggyback without serialization further speeds up the remote invocations with small (10~B) and large (1~MB) objects by 2$\times$ and 2.7$\times$, respectively.}

\subsection{Scalability}
\label{sec:eval_scalability}

We next evaluate the scalability of \SysName{} with regard to internal function invocations and external user requests.

\begin{figure}
    \centering
    \includegraphics[width=.45\textwidth]{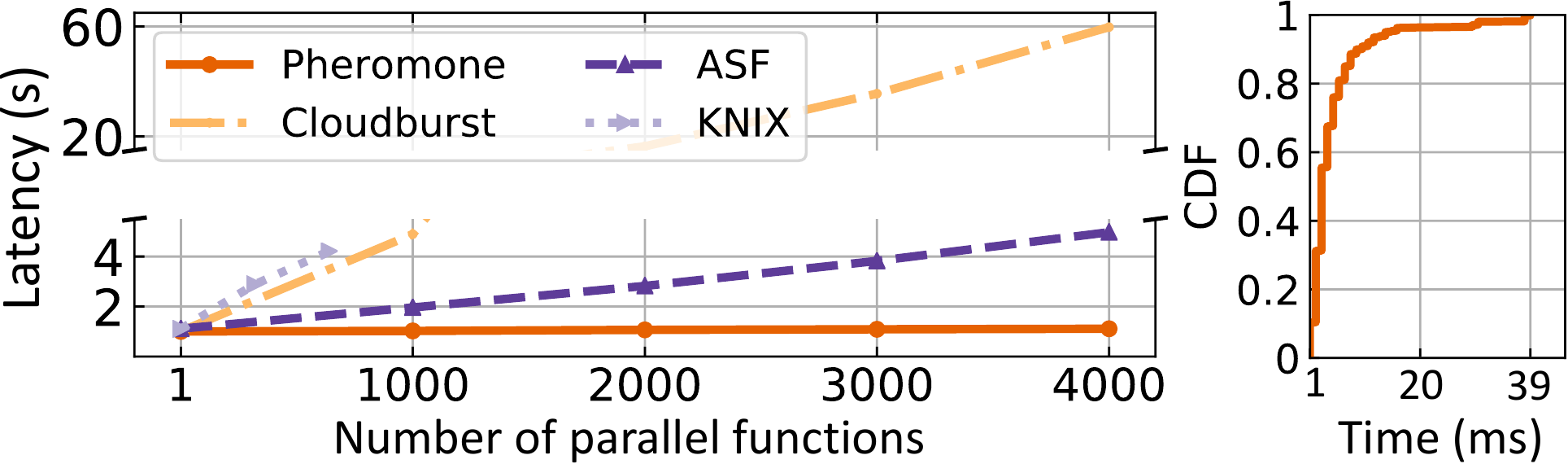}
    \caption{End-to-end latencies with various numbers of parallel functions (left), and the distribution of function start times when executing 4k functions in \SysName (right).}
    \label{fig:parallel_sleep}
    \vspace{-.06in}
\end{figure}

\PHM{Long function chain.}
We start with a long function chain that sequentially invokes a large number of functions~\cite{yu_characterizing_2020}.
Here, each function simply increments its input value by 1 and sends the updated value to the next function, and the final result is the total number of functions.
As shown in Fig.~\ref{fig:function_chain}, we change the number of chained functions, and \SysName{} achieves the best performance at any scale.
Moreover, Cloudburst suffers from poor scalability due to its early-binding scheduling, causing significantly longer latencies when the number of chained functions increases; KNIX cannot host too many function processes in a single container, making it ill-suited for long function chains; ASF incurs the longest latencies due to its high overhead of function interactions.

\PHM{Parallel functions.}
Fig.~\ref{fig:parallel_sleep} (left) evaluates the end-to-end latencies of invoking various numbers of parallel functions, where each function sleeps 1 second.
We run 80 function executors per node in \SysName{} and Cloudburst.
\SysName{} only incurs a negligible latency in large-scale parallel executions, while ASF and Cloudburst incur much higher latencies, e.g., seconds or tens of seconds.
KNIX suffers from severe resource contention when running all workflow functions in the same container, and fails to support highly parallel function executions.
To further illustrate \SysName's behavior in parallel invocations, Fig.~\ref{fig:parallel_sleep} (right) shows the distribution of function start times, where \SysName{} can quickly launch all 4k functions within 40 ms.

\begin{figure}[t!]
    \centering
    \includegraphics[width=.35\textwidth]{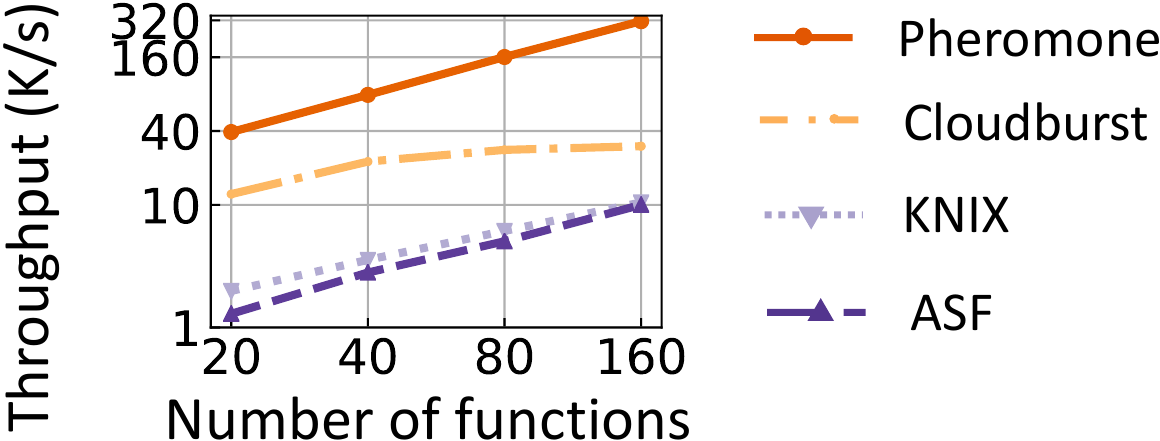}
    \caption{Request throughput when serving requests to no-op functions under various numbers of functions or executors.}
    \label{fig:throughput}
    \vspace{-.1in}
\end{figure}

\PHM{User request throughput.} 
Fig.~\ref{fig:throughput} shows the user request throughput when serving requests to no-op functions using various numbers of executors.
We configure 20 executors on each node in \SysName{} and Cloudburst.
We observe that Cloudburst's schedulers can easily become the bottleneck under a high request rate, making it difficult to fully utilize the executor's resources; KNIX suffers from a similar problem that limits its scalability.
While ASF has no such an issue, it leads to low throughput due to its high invocation overhead (Fig.~\ref{fig:microbench_1}).
Compared with these platforms, \SysName{} ensures better scalability with the highest throughput.

\begin{figure}[t!]
    \centering
    \includegraphics[width=.4\textwidth]{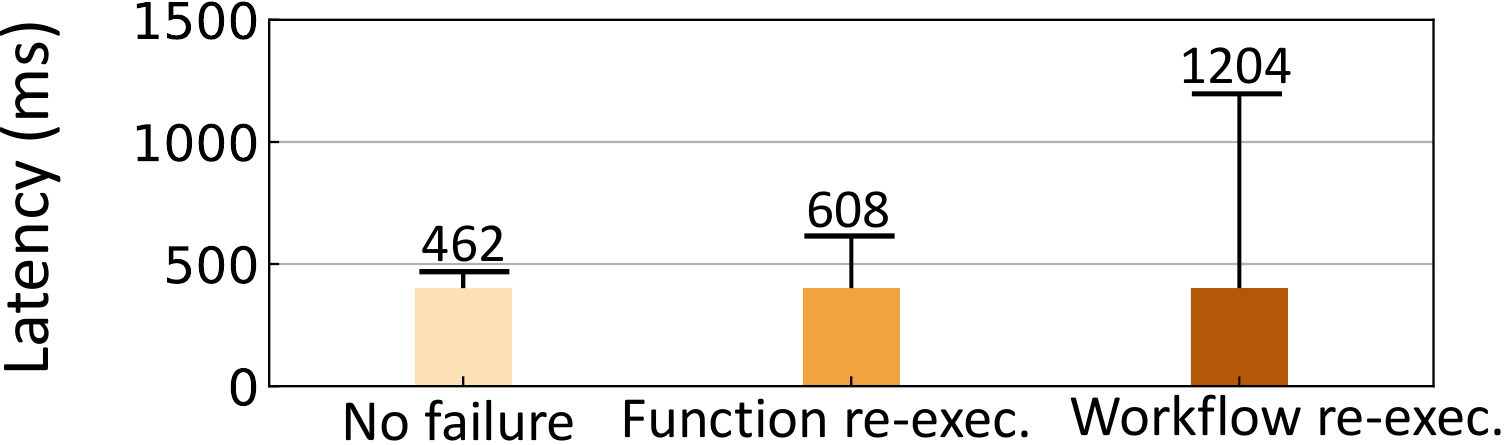}
    \caption{Median and 99$^{th}$ tail latencies of a four-function workflow with no failure, function- and workflow-level re-executions. The numbers indicate the tail latencies.}
    \label{fig:fault_tolerance}
    \vspace{-.1in}
\end{figure}

\subsection{Fault Tolerance}
\label{sec:eval_ft_gc}

\new{We evaluate \SysName's fault tolerance mechanism (\S\ref{sec:system_fault_tolerance}).
We execute a workflow that chains four sleep functions (each sleeps 100 ms), and each running function is configured to crash at a probability of 1\%. 
Fig.~\ref{fig:fault_tolerance} shows the median and 99$^{th}$ tail latencies of the workflow over 100 runs using \SysName's function- and workflow-level re-executions after a configurable timeout.
In particular, the timeout is configured as twice of the normal execution, i.e., 200 ms for each individual function and 800 ms for the workflow.
We also include the normal scenario where no failure occurs. 
Compared with the common practice of workflow re-execution, \SysName's data-centric mechanism allows finer-grained, function-level fault handling, which cuts the tail latency of the workflow from 1204 ms to 608 ms, thus significantly reducing the re-execution overhead.
}

\subsection{Case Studies}
\label{sec:eval_app}

\new{We evaluate two representative applications atop \SysName{}: 
Yahoo's streaming benchmark for advertisement events~\cite{chintapalli_benchmarking_2016}, and a data-intensive MapReduce sort.}

\PHM{\new{Advertisement event stream processing.}}
\new{This application filters incoming advertisement events (e.g., click or purchase), checks which campaign each event belongs to, stores them into storage, and periodically counts the events per campaign every second.
As shown in Fig.~\ref{fig:data_flow_example} (right) and discussed in \S\ref{sec:deploy_app}, the key to enabling this application in serverless platforms is to 
periodically invoke a function to process the events accumulated during the past one second.}

In \SysName, this is straightforward by using the \texttt{ByTime} primitive (\S\ref{sec:orch_api} and Fig.~\ref{fig:script_configure_triggers}).  This application can 
also be implemented easily in DF by using an addressable Entity function for aggregation~\cite{azure_durable_function_aggregator}.
However, it is non-trivial in ASF and we have to resort to a ``serverful'' workaround: one workflow does the regular ``filter-check-store'' for each event and sends the event ID to an external, serverful coordinator; a separate workflow is set up to get triggered every second by the accumulated event IDs sent from the external coordinator, so that it can access and count the associated events per campaign.

\begin{figure}
	\centering
	\includegraphics[width=.45\textwidth]{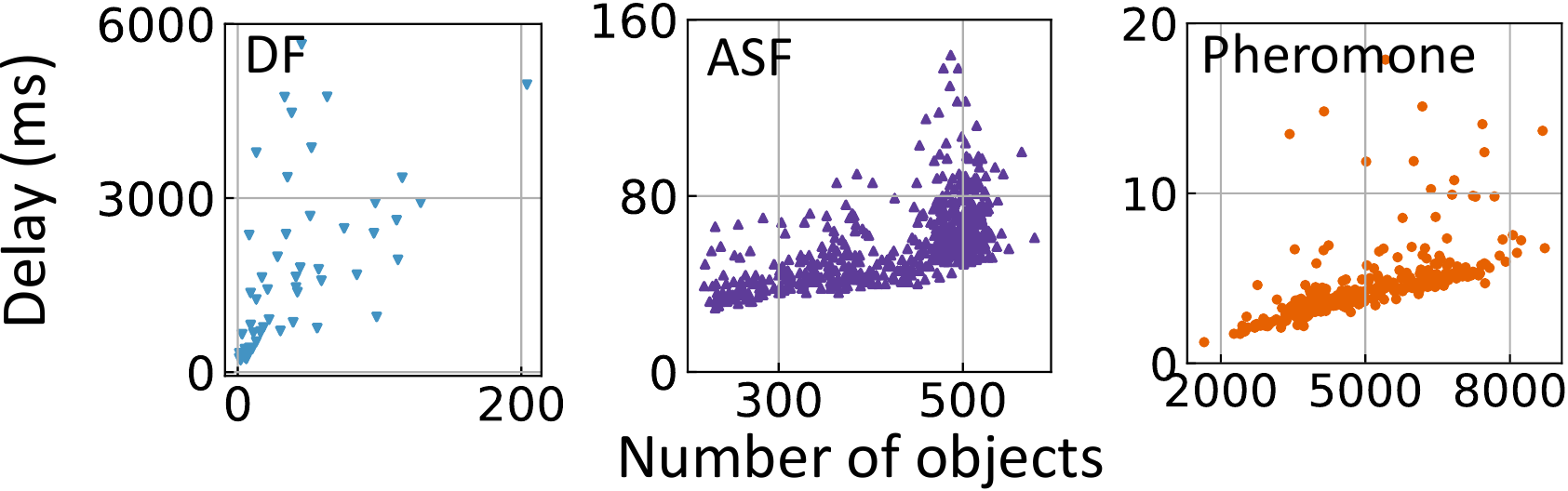}
	\caption{Delays of accessing the accumulated data objects in the advertisement event stream processing. Lower delays and more objects are better. }
	\label{fig:event_object_num}
	\vspace{-.1in}
\end{figure}

\final{Fig.~\ref{fig:event_object_num} compares the performance on \SysName, ASF, and DF. 
We measure the delays of accessing accumulated data objects (i.e., advertisement events), where the lower delays and more objects are better.
For DF, data are not accessed in batches, and thus we measure the queuing delay between the reset request being issued and the Entity function receiving it.
We use up to 40 functions in all these platforms.
DF results in a significant overhead with high and unstable queuing delays, as its Entity function can easily become a bottleneck.
Among the three platforms, \SysName{} performs the best: it can access substantially more accumulated data objects in a much smaller delay.
In summary, \SysName{} not only simplifies the design and deployment for such a stream processing application, but also delivers high performance.}



\PHM{\new{MapReduce sort.}}
We next evaluate how \SysName's data-centric orchestration can easily facilitate MapReduce sort,
a typical data-intensive application. We have built a MapReduce framework atop \SysName{}, called \SysNameMR.
Using the \texttt{DynamicGroup} primitive, \SysNameMR can be implemented in only 500 lines of code, and developers can program standard mapper and reducer~\cite{hadoop} without operating on intermediate data (\S\ref{sec:orch_design}). 
\ifthenelse{\boolean{exclude_appendix}}{}{
Appendix~\ref{sec:appendix_mr} gives an example of writing a MapReduce program in \SysNameMR.}
We compare \SysNameMR with PyWren~\cite{jonas_occupy_2017}, a specialized serverless analytics system built atop AWS Lambda.
Compared with \SysNameMR, PyWren is implemented in about 6k lines of code and supports the map operator only, making it complicated to deploy a MapReduce application:
developers need to explicitly transfer the intermediate data via a shared storage (e.g, Redis) to simulate the reducer, and
carefully configure the storage cluster for improved data exchange. Even with these optimizations, PyWren still suffers from limited performance (and usability).

\begin{figure}
    \centering
    \includegraphics[width=.45\textwidth]{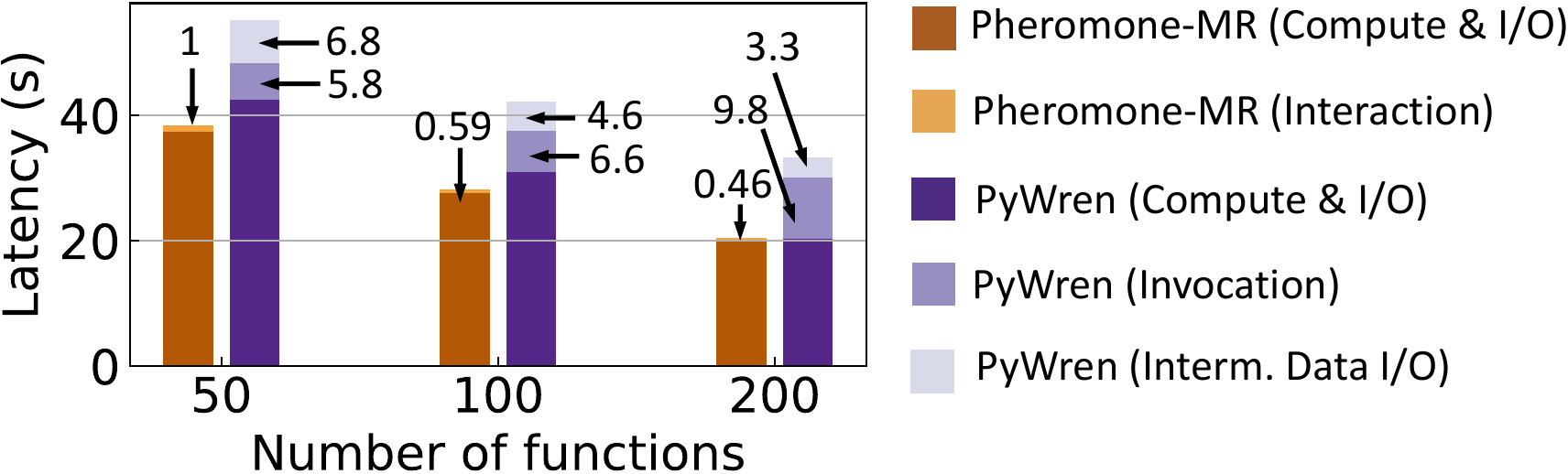}
    \caption{Latencies of sorting 10~GB data using various numbers of functions on PyWren and \SysNameMR. The latency is broken down into: the interaction latency (for PyWren, the invocation and intermediate data I/O), and the latency for compute and I/O. The numbers indicate the former.
    }
    \label{fig:mr_sort}
    \vspace{-.1in}
\end{figure}

\final{We evaluate the performance of \SysNameMR and PyWren with MapReduce sort over 10 GB data, where 10 GB intermediate objects are generated in the shuffle phase.
We allocate 
each \SysName{} executor and each Lambda instance the same resource, e.g., 1 vCPU. We
also configure a Redis cluster for PyWren to enable fast data exchange. 
We measure the end-to-end latencies on \SysNameMR and PyWren using various numbers of
functions, and break down the results into the
function interaction latency and the latency for compute and I/O. The former measures
the latency between the completion of mappers and the start of reducers. For
PyWren, the interaction latency consists of two parts: 1) the invocation latency
of triggering all reducers after mappers return, and 2) the I/O latency of 
sharing intermediate data via Redis. 
As shown in Fig.~\ref{fig:mr_sort}, running more functions in PyWren
improves the I/O of sharing intermediate data, but results in a longer latency in parallel
invocations. Compared with PyWren, \SysNameMR has a significantly lower
interaction latency (e.g., less than 1s), thus improving the end-to-end performance
by up to 1.6$\times$.}

\final{We note that, the limitations of AWS Lambada make PyWren less efficient.
First, because Lambda does not support large-scale map by default~\cite{asf_map}, it needs to implement 
this operation but in an inefficient way which incurs high invocation overheads.
Second, Lambda has a limited support for data sharing, forcing developers to explore an external alternative that incurs high overheads even though using a fast storage (i.e., Redis).
Unlike AWS Lambda, \SysName{} supports rich patterns of function executions while enabling fast data sharing, such that developers can easily build a MapReduce framework and achieve high performance.}


\section{Discussion and Related Work}
\label{sec:discussion}

\PHM{\new{Isolation in \SysName.}}
\new{\SysName provides the container-level isolation between function invocations,
 while functions running on the same worker node share in-memory data objects (\S\ref
 {sec:sys_data_sharing}). Commercial container-based serverless platforms
 often do not co-locate functions from different users to enhance security~\cite{ali_fc}. In this setting, functions on the same worker node can be trusted; hence, it is safe to trade strict isolation for improved I/O performance.
We notice that current serverless platforms have made various trade-offs between performance and isolation. For example, AWS Lambda runs functions in MicroVMs for strong isolation~\cite{agache_firecracker_2020}; KNIX isolates a workflow's functions using processes in the same container for better performance~\cite{akkus_sand:_2018};
recent work proposes lightweight isolation for privacy-preserving serverless applications~\cite{confidential_serverless}. 
\SysName can explore these different trade-offs, which we leave for future work.}

\PHM{\new{Supported languages.}}
\new{\SysName currently supports functions written in C++, but it can be straightforward to support other programming languages.
Specifically, \SysName's executor exposes data trigger APIs (Tabel~\ref{tab:api}) and interacts with other system components, and can serve as a proxy for functions written in different languages.
That being said, \SysName's optimization on fast data exchange via shared memory may not apply to all language runtimes -- only those allowing the direct consumption of byte arrays without (de)serialization, e.g., Python ctype, can benefit from zero-copy data sharing.  The other \SysName designs are still effective regardless of language runtimes.
}

\PHM{Data exchange in serverless platforms.}
Data exchange is a common pain point in today's serverless platforms.
One general approach is to leverage shared storage to enable and optimize data exchange among functions~\cite{carreira_cirrus_2019,fouladi_encoding_nodate,fouladi_laptop_2019,klimovic_pocket:_2018,pu_shuffling_2019,muller_lambada_2020,perron_starling_2020}.  One other approach is to exploit data locality to improve performance, e.g., by placing workflow functions on a single machine~\cite{sreekanti_cloudburst_2020,shillaker_faasm_2020,jia_nightcore_2021,mahgoub_sonic_nodate,kotni_faastlane_nodate,tang_lambdata_2020}.
Moreover, OFC~\cite{mvondo_ofc_2021} and Faa\$T~\cite{romero_faat_2021} provide the autoscaling cache for individual applications.
Shredder~\cite{zhang_narrowing_2019} and Zion~\cite{sampe_data-driven_2017} push the function code into storage.
Wukong~\cite{carver_wukong_2020} enhances the locality of DAG-based parallel workloads at the application level. 
Lambdata~\cite{tang_lambdata_2020} makes the intent of a function's input and output explicit for improved locality; however, it does not provide a unified programming interface for expressive and simplified function interactions, and its performance is heavily bound to Apache OpenWhisk~\cite{openwhisk, kotni_faastlane_nodate}.
\section{Conclusion}


This paper revisits the function orchestration in serverless computing, and advocates a new design paradigm that a serverless platform needs to:
1) break the tight coupling between function flows and data flows, 
2) allow fine-grained data exchange between functions of a workflow, and 
3) provide a unified and efficient approach for both function invocations and data exchange.  
With this data-centric paradigm, we have designed and developed \SysName, a new serverless platform which achieves all the desired properties, namely, rich expressiveness, high usability, and wide applicability. \SysName is open-sourced, and outperforms current commercial and open-source serverless platforms by orders of magnitude in terms of the latencies of function invocation and data exchange.


\section*{Acknowledgments}

We thank the anonymous reviewers and our shepherd, Yiying Zhang, for their
insightful comments that helped shape the final version of this work. We also
thank Yuheng Zhao for his help in experiments, and Chenggang Wu for his
valuable feedback at the early stage of this work. This work was supported in
part by RGC GRF Grants 16202121 and 16210822. Minchen Yu was supported in
part by the Huawei PhD Fellowship Scheme.

\ifthenelse{\boolean{exclude_appendix}}{}{
\appendix
\section{Appendix}
\label{sec:appendix}

\setcounter{figure}{0}
\renewcommand\thefigure{A.\arabic{figure}} 


  
  

\subsection{Trigger Example}
\label{sec:appendix_trigger_interface}

\SysName provides a common interface for developers to implement customized
 trigger primitives. There are three
 main methods of the interface (\S\ref{sec:orch_design}, Fig.~\ref{fig:script_trigger_interface}). 
 Fig.~\ref{fig:script_by_batch_size} provides an example of using the interface to implement the \texttt{ByBatchSize} trigger primitive.

\begin{figure}[t!]
\begin{lstlisting}[style=mycpp, escapechar=|, escapeinside={(*}{*)}]
(*s: the batch size specified by developers*)
(*t: the target function of this trigger*)
(*B: a buffer of objects for next invocation*)
(*L: a set of statuses of running source functions*)

(*action\_for\_new\_object(obj)\{*)
  A (*$\gets$*) {}
  (*remove the source function of obj from L*)
  B.add(obj)
  if (B.size() >= s){
    (*a $\gets$ trigger action for function t with objects in B*)
    A.add(a)
    B.clear()
  }
  return A
}

(*notify\_source\_func(info of source function)\{*)
  (*add relevant info of source function to L*)
}

(*action\_for\_rerun(session)\{*)
  A (*$\gets$*) {}
  if ((*L contains functions in this session*)){
    (*A $\gets$ trigger actions for relevant functions in L*)
  }
  return A
}
\end{lstlisting}
\vspace{-.4in}
\caption{A pseudocode implementation of \texttt{ByBatchSize} primitive using the provided trigger interfaces.}
\label{fig:script_by_batch_size}
\end{figure}

\subsection{Advertisement Event Stream}
\label{sec:appendix_aes}

Fig.~\ref{fig:script_full_stream_script} shows the script of deploying the
event stream processing application (case study~\#1 in \S\ref
{sec:eval_app}) in \SysName. The application has three functions,
\texttt{preprocess}, \texttt{query\_event\_info}, and \texttt{aggregate},
chained through two buckets (see Fig.~\ref{fig:pri_all}, right). The
script first registers the application with the name and three
functions (lines 1-3), and then configures two bucket triggers, \texttt
{Immediate} and \texttt{ByTime}, to invoke functions \texttt{query\_event\_info}
(lines 6-11) and \texttt{aggregate} (lines 14-20), respectively. Fig.~\ref
{fig:script_stream_cpp_snippet} shows the code snippets of how the two producer 
functions \texttt{preprocess} and \texttt{query\_event\_info}
interact with the two buckets. Each function creates an empty object by specifying its
target bucket and key name (lines 2 and 7). Once the object value is
computed (omitted in the code snippet) and the object is ready to be consumed, the
function sends it to the bucket via the \texttt{send\_object} method, thereby triggering the
next function. 
Pheromone also provides a function-oriented interface for application deployment, which we detail in Appendix~\ref{sec:foi}.

\begin{figure}[t!]
\begin{lstlisting}[style=mypython, escapechar=|]
app_name = 'event-stream-processing'
functions = ['preprocess','query_event_info','aggregate']
client.register_app(app_name, functions)

# configure the first bucket trigger.
bck_name = 'immed_bck'
trig_name = 'immediate_trigger'
prim_meta = {'function':'query_event_info'}
client.create_bucket(app_name, bck_name)
client.add_trigger(app_name, bck_name, trig_name, \ 
			IMMEDIATE, prim_meta)

# configure the second bucket trigger.
bck_name = 'by_time_bck'
trig_name = 'by_time_trigger'
prim_meta = {'function':'aggregate', 'time_window':1000}
re_exec_rules = ([('query_event_info', EVERY_OBJ)], 100)
client.create_bucket(app_name, bck_name)
client.add_trigger(app_name, bck_name, trig_name, \
      BY_TIME, prim_meta, hints=re_exec_rules)
\end{lstlisting}
\vspace{-.4in}
\caption{Deploying the stream processing workflow.}
\label{fig:script_full_stream_script}
\end{figure}

\begin{figure}[t!]
\begin{lstlisting}[style=mycpp, escapechar=|]
// specify the bucket in function preprocess
obj = library->create_object("immed_bck", key_name);
... // operate on obj
library->send_object(obj);

// specify the bucket in function query_event_info
obj = library->create_object("by_time_bck", key_name);
... // operate on obj
library->send_object(obj);
\end{lstlisting} 
\vspace{-.4in}
\caption{Code snippets of interacting with buckets.}
\label{fig:script_stream_cpp_snippet}
\end{figure}


\subsection{MapReduce atop \SysName}
\label{sec:appendix_mr}

\SysName's data-centric function orchestration provides a natural support for
 implementing data-flow applications in the serverless cloud. We have
 developed a MapReduce framework atop \SysName using the system-provided
\texttt{DynamicGroup} primitive and other APIs, which we call \texttt{\SysNameMR}. 
In particular, \SysNameMR configures \texttt{DynamicGroup} bucket triggers for a MapReduce program. It can accept the code snippets of mapper and reducer from clients, and package them into two Pheromone functions, i.e., map and reduce.
The Pheromone map functions execute the user-provided code, and then automatically send the intermediate objects to the pre-configured bucket and specify which data group each object belongs to.
When map functions all complete, reduce functions get invoked by the bucket trigger to consume intermediate objects and output the final results. 
Fig.~\ref{fig:script_mapreduce} shows a classical WordCount example 
in \texttt{\SysName{}-MR}. 
Note that our implementation of \texttt{\SysNameMR}
includes only 380 lines of C++ function wrappers and 120 lines of Python interfaces.
In comparison, PyWren~\cite{jonas_occupy_2017}, an equivalent system built on Lambda, 
requires about 6k lines of Python for executing map-only operations.



\begin{figure}[t!]
\begin{lstlisting}[style=mycpp, escapechar=|]
void map_function(const char* input, size_t data_size) {
    string input_str{input, data_size};
    vector<string> words;
    split_string(input_str, ' ', words);
    for (auto &w : words) emit(w, 1);
}

void reduce_function(string key, vector<int> &values) {
    int sum = 0;
    for (auto& v : values) sum += v;
    emit(key, sum);
}
\end{lstlisting}
\vspace{-.4in}
\caption{A WordCount example in \SysNameMR. Users simply write map (line 1-6) and reduce (line 8-12) functions.}
\label{fig:script_mapreduce}
\end{figure}

\begin{figure}[t!]
\begin{lstlisting}[style=mypython, escapechar=|]
app_name = 'event-stream-processing'
functions = ['preprocess','query_event_info','aggregate']
# dependency: (source_functions, target_functions, type)
dep1 = (['preprocess'],['query_event_info'],DIRECT)
# Include extra metadata (1000 ms timeout) in dependency
dep2 = (['query_event_info'],['aggregate'],PERIODIC,1000)
client.register_app(app_name, functions, [dep1, dep2])
# The two functions preprocess and query_event_info can
# directly create objects w/o specifying the bucket name:
#     library->create_object();
\end{lstlisting}
\vspace{-.4in}
\caption{Deploying the stream processing workflow using function-oriented interface.}
\label{fig:script_stream_function_oriented}
\end{figure}

\subsection{Function-Oriented Interface}
\label{sec:foi}
\SysName{} also provides a function-oriented interface that encapsulates the
 data trigger APIs. The interface hides the detailed data consumption and
 allows developers to directly express function interactions without
 specifying buckets for workflow functions. Fig.~\ref
 {fig:script_stream_function_oriented} gives an example of using this
 interface to deploy the aforementioned event stream processing workflow
 (case study~\#1 in \S\ref{sec:eval_app}). Developers simply specify the
 dependencies between functions (lines 4 and 6), which can be interpreted as
 data triggers. That is, the \texttt{DIRECT} and \texttt
 {PERIODIC} dependencies are translated into the \texttt
 {Immediate} and \texttt{ByTime} trigger primitives (Fig.~\ref
 {fig:script_full_stream_script}), respectively. It is worth mentioning that
 the function-oriented interface cannot support complex workflows that need
 the function-side knowledge to perform fine-grained data exchange
 (e.g., MapReduce), for which data trigger APIs are needed.

\subsection{Evaluation of Garbage Collection}
\label{sec:eval_gc}

\SysName garbage-collects in-memory data objects after the associated invocation request has been fully served (\S\ref{sec:sys_data_sharing}). 
We evaluate the effectiveness of garbage collection (GC) with a synthesized workflow.
The workflow contains two chained functions, which generates 1~MB dummy data and 
keeps alive for a random period of time (e.g., sleep). The keep-alive period ranges from 1~ms to 1000~ms and is sampled following the log-normal distribution, a typical distribution of function execution time in a production platform~\cite{shahrad_serverless_2020}. We concurrently run this workflow in \SysName to fully occupy the executor resources, and measure
the memory usage over time with and without GC.
As shown in Fig.~\ref{fig:shm_gc}, enabling GC in \SysName can timely 
evict unused data objects from the in-memory store, significantly reducing the 
memory footprint.

\begin{figure}
    \includegraphics[width=.47\textwidth]{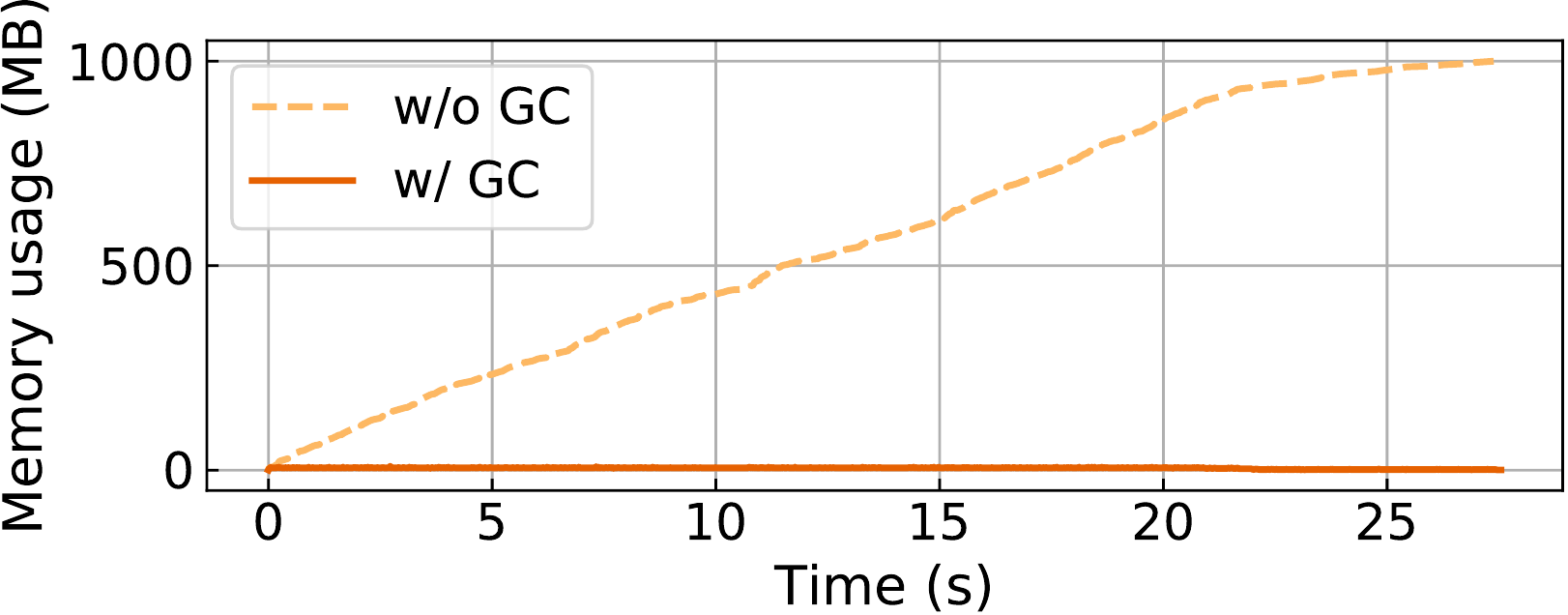}
    \caption{Memory usage of shared-memory object store with and without garbage collection.}
    \label{fig:shm_gc}
    \vspace{-.1in}
\end{figure}
} 


\bibliographystyle{plain}
\bibliography{references}

\end{document}